\documentclass[iop]{emulateapj}
\usepackage{epstopdf}
\usepackage{apjfonts,graphicx}
\usepackage{epsf}
\usepackage{hyperref}
\usepackage{url}
\usepackage[flushleft]{threeparttable}
\usepackage{mathtools}

\begin{document}

\slugcomment{To be submitted to ApJ}
\shortauthors{Li et al.}
\shorttitle{Star Cluster Formation in Cosmological Simulations}

\title{Star cluster formation in cosmological simulations. I. properties of young clusters}
        
\author{Hui Li\altaffilmark{1*}, Oleg Y. Gnedin\altaffilmark{1}, Nickolay Y. Gnedin\altaffilmark{2,3,4}, Xi Meng\altaffilmark{1,5}, Vadim A. Semenov\altaffilmark{3,4}, Andrey V. Kravtsov\altaffilmark{3,4,6}}

\altaffiltext{1}{Department of Astronomy, University of Michigan, Ann Arbor, MI 48109, USA}
\altaffiltext{2}{Particle Astrophysics Center, Fermi National Accelerator Laboratory, Batavia, IL 60510, USA}
\altaffiltext{3}{Kavli Institute for Cosmological Physics, University of Chicago, Chicago, IL 60637, USA}
\altaffiltext{4}{Department of Astronomy \& Astrophysics, University of Chicago, Chicago, IL 60637, USA}
\altaffiltext{5}{Department of Astronomy, Peking University, 100871, Beijing, China}
\altaffiltext{6}{Enrico Fermi Institute, The University of Chicago, Chicago, IL 60637, USA}
\altaffiltext{*}{hliastro@umich.edu}

\date{\today}

\begin{abstract}
We present a new implementation of star formation in cosmological simulations, by considering star clusters as a unit of star formation. Cluster particles grow in mass over several million years at the rate determined by local gas properties, with high time resolution. The particle growth is terminated by its own energy and momentum feedback on the interstellar medium. We test this implementation for Milky Way-sized galaxies at high redshift, by comparing the properties of model clusters with observations of young star clusters. We find that the cluster initial mass function is best described by a Schechter function rather than a single power law. In agreement with observations, at low masses the logarithmic slope is $\alpha\approx 1.8-2$, while the cutoff at high mass scales with the star formation rate. A related trend is a positive correlation between the surface density of star formation rate and fraction of stars contained in massive clusters. Both trends indicate that the formation of massive star clusters is preferred during bursts of star formation. These bursts are often associated with major merger events. We also find that the median timescale for cluster formation ranges from 0.5 to 4 Myr and decreases systematically with increasing star formation efficiency. Local variations in the gas density and cluster accretion rate naturally lead to the scatter of the overall formation efficiency by an order of magnitude, even when the instantaneous efficiency is kept constant. Comparison of the formation timescale with the observed age spread of young star clusters provides an additional important constraint on the modeling of star formation and feedback schemes.
\end{abstract}

\keywords{galaxies: formation --- galaxies: star clusters ---  globular clusters: general}

\section{Introduction}
One of the most important problems in astrophysics is understanding the formation and evolution of galaxies. Although the $\Lambda$ cold dark matter paradigm shows great success in reproducing observations on large scales and builds a solid theoretical framework \citep{planck_etal14, vikhlinin_etal09, springel_etal06}, many galaxy formation questions, especially those that involve baryonic physics, are not yet answered.

A self-consistent way of modeling the baryonic component of galaxies is to run simulations that include all relevant physics such as gravity, hydrodynamics, star formation, radiation transport, etc. This approach has already proved to be a very important tool. Current simulations are making rapid progress in reproducing different types of galaxies and global scaling relations, such as the Kennicutt-Schmidt relation and star formation history, by realizing the importance of stellar feedback in shaping the properties of interstellar medium \citep[ISM; for a recent review, see][]{somerville_dave15}. Various stellar feedback mechanisms suppress star formation activity at high redshifts and retain a gas reservoir for star formation at low redshifts.  This feedback loop allows the self-regulation of star formation in galaxies.

Although many processes of stellar feedback have been explored \citep[e.g.][]{katz92, navarro_white_93, stinson_etal06, governato_etal07, governato_etal12, scannapieco_etal08, hopkins_etal11,hummels_bryan12, booth_etal13, agertz_etal13, stinson_etal13, ceverino_etal14, salem_bryan_14, hopkins_etal14, keller_etal15}, the star formation prescription has not changed for over two decades since \citet{katz92} and \citet{cen_ostriker_92}. In most of the current cosmological simulations, star particles are formed in cold and dense gas, with the rate that is calculated by assuming a fixed efficiency per free-fall time, $\epsilon_{\rm ff}$. One exception is the recent implementation of the turbulence-based $\epsilon_{\rm ff}$ by \citet{semenov_etal15}. Star particles are created by a Poisson process, with their masses set beforehand and therefore, unrelated to the feedback they produce. Yet, the efficiencies of star formation and feedback are closely interrelated, as both affect key properties of galaxies \citep{agertz_kravtsov_15, agertz_kravtsov_16}, and thus must be treated self-consistently.

One possible solution is to grow stellar particles over time and let their own feedback terminate star formation locally.  This causes a number of issues that need to be resolved: How to choose the size of a star-forming region?  How to treat interactions of neighboring regions?  How to tell if the continuous model is more realistic than the instantaneous particle creation?  Important constraints on these come from observations of clustering of young stars.  

Most stars form in clusters and associations, which can be considered as building blocks of the stellar component of galaxies \citep[e.g.][]{lada_lada03}.  Star clusters follow a well-defined initial mass function (CIMF), analogously to the mass function of individual stars.  It follows an approximate power law at low masses but has a steeper falloff at high mass, which can be described by the Schechter function \citep[e.g.][]{portegies_zwart_etal10}.  The maximum cluster mass scales with the star formation rate of its host galaxy.  Clusters form within the dense parts (cores and clumps) of giant molecular clouds (GMCs).  Stars in the youngest clusters, still partially embedded in the molecular gas, show a spread of ages of only a few Myr \citep{maclow_klessen_04, hartmann_etal12, hollyhead_etal15}. Even though resolving the sizes of young clusters ($\sim 1$~pc) is still beyond capabilities of current cosmological simulations, resolving GMC structure is within reach \citep[e.g.][]{hopkins_etal14, read_etal15, wetzel_etal16}. Therefore, the implementation of star formation can be made more realistic if stellar particles corresponded directly to individual clusters forming within GMCs.  The observed properties of young star clusters, such as the CIMF and the cluster formation timescale, can serve as tests of the star formation and feedback prescriptions on a scale much smaller than what is typically used in galaxy formation simulations ($\sim 1$~kpc).

In this paper, we develop a new model for implementing star formation by considering star cluster as a unit of star formation. In contrast to all other cosmological simulations where stellar particles are formed instantaneously, the formation of a stellar particle in our model is resolved in time over a period of a few Myr and terminated by its own feedback. The final cluster masses are set self-consistently, given the implemented physics of stellar feedback. We show that the shape of the model CIMF is consistent with observations of young star clusters, but the cluster formation timescale decreases systematically with increasing local efficiency of star formation and strength of stellar feedback.

We also show that this model leads to the scatter of the cluster formation efficiency by an order of magnitude, even when the instantaneous efficiency per free-fall time is kept constant.  This scatter is caused by local variations in the gas density and cluster accretion rate, which are affected by the cluster feedback.  The variation of the efficiency is predicted by analytical models of star formation in supersonic turbulent GMCs \citep[e.g.][]{krumholz_mckee05, padoan_nordlund11, federrath_klessen12, hennebelle_chabrier13}, although detailed comparison with our results is beyond the scope of this paper.  Small-scale simulations of turbulent cascade also predict systematic variation of $\epsilon_{\rm ff}$ with the virial parameter \citep[e.g.,][]{padoan_etal12, kritsuk_etal13, federrath15}. Observational constraints on $\epsilon_{\rm ff}$ on the GMC scale vary from $\sim 0.01\%$ to $\sim 10\%$ \citep[e.g.,][]{murray11, evans_etal14, vutisalchavakul_etal16}. Although estimates for some clouds reach values as high as $\epsilon_{\rm ff}\sim 30\%$, these values  may be artificially inflated by different evolutionary phases of the GMC \citep{feldmann_gnedin11}.  The fact that our implementation naturally leads to scatter of cluster formation efficiency paves the way towards more realistic modeling of star formation in the turbulent galactic ISM.

In Section~\ref{sec:setup}, we describe the simulation setup and detailed implementation of this sub-grid model. In Section~\ref{sec:result}, we examine the global star formation history of the simulated galaxies. In Section~\ref{sec:CIMF}, we investigate the shape of the cluster initial mass function and its environmental dependence, and compare them with observations of young massive star clusters in nearby galaxies. In Section~\ref{sec:timescale}, we demonstrate an important effect of the star formation efficiency on the cluster formation timescale. In Section~\ref{sec:discussion}, we discuss the origin of CIMF in the turbulent ISM and compare our implementation with previous models. Finally, we summarize our conclusions in Section~\ref{sec:sum}.

\section{Simulation Setup}\label{sec:setup}

The simulations were run with the Eulerian gasdynamics and N-body Adaptive Refinement Tree (ART) code \citep{kravtsov_etal97, kravtsov99, kravtsov03, rudd_etal08}. The latest version of ART includes several new physical ingredients that make it suitable for investigating the detailed star formation processes in a cosmological context. It includes an updated version of three-dimensional radiative transfer of ionizing and ultraviolet radiation using the Optically Thin Variable Eddington Tensor approximation \citep{gnedin_abel01}. Both the local ionizing radiation from star particles \citep{gnedin14} and the extra-galactic background \citep{haardt_madau01} are considered as the ionization sources that feed into the radiative transfer solver. The ionization states of various species of hydrogen (HI, HII, $\rm H_2$) and helium (HeI, HeII, HeIII), as well as the cooling and heating rates, are calculated based on the non-equilibrium chemical network. Finally, a phenomenological model of $\rm H_2$ formation and self-shielding on dust grains \citep{gnedin_kravtsov11} allows us perform a more realistic modeling of star formation based on local molecular component. Recently a subgrid-scale (SGS) model for the numerically unresolved turbulence was implemented in the ART code. Turbulence, generated by gravitational instabilities as well as kinetic feedback from stars and AGN, is an important ingredient for star formation \citep{mckee_ostriker07}. This SGS model has been tested in isolated disk simulations \citep{semenov_etal15} and in this paper it is applied to cosmological simulations.

We run cosmological simulations in a periodic box of size $L_{\rm box}=4$ Mpc comoving. The initial condition is selected and tested by collisionless runs so that the central galaxy has total mass $M_{200}\approx10^{12}M_\sun$ at $z=0$. There are also a few satellite galaxies with halo masses in the range $10^{10}-10^{11}M_\sun$.

This initial condition has a non-zero ``DC mode'' that corrects the deviation of the cosmological evolution due to the difference between the average matter density in the box and the average matter density in the whole universe \citep{gnedin_etal11}. A constant parameter $\Delta_{\rm DC}$, which represents the density fluctuation level of the current box, determines the relationship between the expansion rate of the box and the expansion rate of the universe:
\begin{equation}
a_{\rm box} = \frac{a_{\rm uni}}{[1+\Delta_{\rm DC}D_+(a_{\rm uni})]^{1/3}},
\end{equation}
where $a_{\rm box}$ and $a_{\rm uni}$ are the scale factors of the simulation box and the global universe, respectively. $D_+(a)$ is the linear growth factor of density perturbation at scale factor $a$. Our initial condition has $\Delta_{\rm DC}=-1.02$, that is, a slightly underdense region. 

The ART code uses adaptive mesh refinement, which increases spatial resolution in high density regions during simulation runtime.  All of our simulations start with a $128^3$ root grid, which gives the mass of dark matter particle $m_{\rm DM}=1.05\times10^6M_\sun$ and size of the root cell 31.25 kpc comoving. We allow a maximum of ten additional refinement levels, which gives us spatial resolution of $L_{10}=4\times10^6/128/2^{10}\approx30$ pc comoving. At $z\approx4$, the physical size of a $10^{\rm th}$ level cell is about 6 pc, smaller than a typical size of a giant molecular cloud. We employ a combination of the Lagrangian refinement criteria (both DM and gas mass) and the Jeans refinement criteria in the simulation. The cell will be refined if either criteria is fulfilled. For the Lagrangian refinement criteria, the cell will be refined if either the DM mass of the cell exceeds $3 f_{\rm tol} m_{\rm DM}\Omega_{\rm m}/\Omega_{\rm DM}$, or the gas mass exceeds $3 f_{\rm tol} m_{\rm DM}\Omega_{\rm b}/\Omega_{\rm DM}$, where the refinement split tolerance $f_{\rm tol}=0.6$. For the Jeans refinement criteria, cells larger than twice the local Jeans length $\lambda_{\rm Jeans}$ will be refined. The local Jeans length $\lambda_{\rm Jeans}$ is defined as
\begin{equation}
\lambda_{\rm Jeans}=v_{\rm gas}\tau_{\rm ff} \approx \sqrt{\frac{\pi(c_s^2+\sigma_v^2)}{G\rho_{\rm gas}}},
\end{equation}
where $c_s$ is the sound speed of the cell, and $\sigma_v$ is the local gas velocity dispersion computed by the root mean square value of the velocity differences between a given cell and its six immediate neighbors.

We adopt a $\Lambda$CDM cosmology with $\Omega_{\rm m}=0.304$, $\Omega_{\rm b}=0.048$, $h=0.681$, and $\sigma_8=0.829$ that is consistent with the most recent $\textit{Planck}$ result \citep{planck_etal15}.

\begin{table*}
\begin{center}
  \begin{threeparttable}
\caption{\sc Model Parameters}
\label{tab:par}
\def\arraystretch{1.5}
\begin{tabular}{cccccccc}
\tableline
\multicolumn{1}{c}{Models} &
\multicolumn{1}{c}{SFE10 (fiducial)} &
\multicolumn{1}{c}{SFE20} &
\multicolumn{1}{c}{SFE100} &
\multicolumn{1}{c}{LOWRHO} &
\multicolumn{1}{c}{TURB50} &
\multicolumn{1}{c}{TURBSF} &
\multicolumn{1}{c}{TURBSF2} \\
\tableline
$\rho_{\rm crit}  ({\rm cm}^{-3})$ &  1000 & 1000 & 1000 &  10 & 1000	& - &-  \\
$T_{\rm crit} ({\rm K})$ &  20000 & 20000 & 20000 & 20000 & 20000	& - & -  \\
$f_{\rm H2, crit}$ &  0.5 & 0.5 & 0.5 & 0.5 & 0.5	 & - & 0.5 \\
$\epsilon_{\rm ff}$ &  0.1  & 0.2 & 1.0 & 0.1 & 0.1 & - & - \\
$f_{\rm turb}$ &  0.1 & 0.1 & 0.1  &  0.1 & 0.5	& 0.5 & 0.5\\
\tableline
\end{tabular}
    \begin{tablenotes}
      \small
      \item \textbf{Note.} Other fixed parameters for all models: $\tau_{\rm max}=15$ Myr; $M_{\rm th}=10^3M_\sun$; $D_{\rm GMC}=10$ pc; feedback speed ceiling $v_{\rm fb}^{\rm ceiling}=5000$ km/s; feedback temperature ceiling $T_{\rm fb}^{\rm ceiling}=10^8$ K.
    \end{tablenotes}
  \end{threeparttable}
\end{center}
\end{table*}

\subsection{Continuous Cluster Formation}\label{sec:CCF}

We introduce a new model for the formation of stellar particles in cosmological simulations: continuous cluster formation. In this prescription, cluster particles are formed within a spherical region of fixed physical size, $D_{\rm GMC}\sim10$ pc. This setup can avoid sudden changes of cell size by mesh refinement and gradual changes due to the expansion of the Universe.

In most of our runs, cluster particles are created only in cold ($T<T_{\rm crit}$), dense ($\rho>\rho_{\rm crit}$) cells with high molecular fraction ($f_{\rm H2}>f_{\rm H2, crit}$) \footnote{In contrast to the fixed $\epsilon_{\rm ff}$ runs, the turbulence-based $\epsilon_{\rm ff}$ runs does not require an ad hoc density, temperature, or molecular fraction threshold for star formation, as $\epsilon_{\rm ff}$ is naturally suppressed in warm diffuse gas \citep[see][]{semenov_etal15} .}. Also, we allow particle creation only when the spherical cluster formation region is located at the local density maximum by comparing the density of a given cell with its 6 immediate neighbors. Once created, cluster particles are labeled ``active'' and will grow during the cluster formation timescale $\tau_{\rm max}=15$ Myr. Clusters with age older than $\tau_{\rm max}$ will be labeled ``inactive'' and stop growing, because the observed age spread in young star clusters is typically below 15 Myr. Another constraint on cluster formation is that new cluster particle can be created only in a cell that does not already contain another active particle. Otherwise, we will grow the existing cluster instead of creating a new one. 

All cluster particles in our simulations are considered to be collisionless elements, which go into the gravity solver. The Poisson equation is solved by FFT at the root level, and by the relaxation method at all refinement levels. These particles are mapped to Eulerian grids; therefore, they experience the same potential irrespective of their masses.

Since the resolution of the simulations can reach 30 comoving pc, a sphere of physical diameter 10 pc can be larger than a single cell at high redshifts. In this case, the neighboring cells that are covered by the sphere should also participate in cluster formation. We developed an algorithm that returns gas properties of central, as well as 26 neighboring cells, and calculates the total star formation rate (SFR) within the sphere. But the volume participating in cluster formation cannot be larger than the 27 cells.

The particle growth rate is calculated as follows. The SFR density of the central cell is:
\begin{equation} \label{eq:sfr_c}
\dot{\rho}_* = \epsilon_{\rm ff} \frac{ f_{\rm H_2} \rho} {\tau_{\rm ff}},
\end{equation}
where $\tau_{\rm ff}=\sqrt{3\pi/32G\rho_c}$ is the free-fall time of the central cell,  $\epsilon_{\rm ff}$ is the star formation efficiency per free-fall time, and $f_{\rm H_2}$ is calculated based on the non-equilibrium chemical network that is described in \citep{gnedin_kravtsov11}. The total mass accumulation rate within the sphere is the sum of the contributions from the central as well as 26 neighboring cells:
\begin{equation} \label{eq:cfr}
\dot{M}=\sum_i f_{{\rm sp},i} V_i \dot{\rho}_{*,i}= \frac{\epsilon_{\rm ff}}{\tau_{\rm ff}} \sum_i  f_{{\rm sp}, i} V_i f_{{\rm H_2}, i} \rho_{{\rm gas}, i},
\end{equation}
where $V_i$ is the volume of cell $i$, and $f_{{\rm sp}, i}$ is the fraction of $V_i$ that is located within sphere. Here the SFR density of the neighboring cells is estimated by the free-fall time of the central cell, since the collapse of the whole spherical cluster formation region is dominated by the central cell. The mass increment during simulation timestep $\Delta t$ is then $\dot{M}\Delta t$.

The star formation efficiency $\epsilon_{\rm ff}$ in model SFE10 (fiducial) is set to be ten percent. Model SFE20 ($\epsilon_{\rm ff}=20\%$) and SFE100 ($\epsilon_{\rm ff}=100\%$) test the influence of the star formation efficiency on the global SFR as well as cluster particle properties. Recently, \citet{padoan_etal12} found a relationship between $\epsilon_{\rm ff}$ and cloud virial parameter ($\alpha_{\rm vir}$) by analyzing MHD simulations of turbulent GMCs. \citet{semenov_etal15} implemented a SGS turbulence model in the ART code to study the influence of turbulence on gas dynamics and explore the turbulence-based star formation efficiency in isolated disk simulations. In this paper, we setup a simulation TURBSF to test this model in cosmological runs. The SFR density is evaluated as $\dot{\rho}_* = \epsilon_{\rm ff}(\alpha_{\rm vir}) \rho / \tau_{\rm ff}$. We also apply this turbulence-based $\epsilon_{\rm ff}$ on molecular gas (TURBSF2) by adding the molecular fraction $f_{\rm H_2}$ in the above rate: $\dot{\rho}_* = \epsilon_{\rm ff}(\alpha_{\rm vir})  f_{\rm H_2} \rho / \tau_{\rm ff}$. The parameters of the runs are listed in Table \ref{tab:par}.

For all simulations discussed in this paper, we take 10 pc, a typical size of dense clumps in GMCs \citep[e.g.,][]{murray11}, as the diameter of the cluster formation sphere, $D_{\rm GMC}$. This value is consistent with the size of massive cluster-forming clouds that are observed in the recent sub-mm ATLASGAL survey \citep{urquhart_etal14}. The range of the effective radius for massive proto-cluster candidates in that survey is about 5-10 pc. The value of $D_{\rm GMC}$ is tightly constrained. Any small sphere size would not contain enough material to form massive clusters, while a significantly larger sphere size could include several regions that should be forming separate clusters.

Because $D_{\rm GMC}$ is an important parameter in our model, we test its influence on the formation of cluster particles by varying its value around 10 pc. In the case of $D_{\rm GMC}=5$ pc, the sphere is small enough so that it completely resides within the central cell and cannot reach any neighboring cells, for all redshifts of interest $z<11$. We find that the masses of the model clusters are smaller than those in the case of $D_{\rm GMC}=10$ pc by a factor of $1.8-2.0$. The CIMF exhibits a systematic shift to lower masses so that clusters with mass larger than several million solar mass are seldom formed. We have done additional test runs with $D_{\rm GMC}$ increasing from 5 to 15 pc, and found that the CIMF first extends to higher masses, and then gradually saturates at $D_{\rm GMC}>10$ pc. This suggests that a sphere with $D_{\rm GMC}=10$ pc contains majority of the molecular gas that will collapse into a single cluster-forming region.

However, when the size of the sphere is much larger than the gas cell, $D_{\rm GMC}>3L_{\rm cell}$, the sphere will cover more than one layer of neighboring cells. Extracting gas properties of all cells within the sphere makes the book-keeping complicated and significantly reduces the computation efficiency. Therefore, we only allow cluster particles to grow their mass from the gas within the closest 27 cells. With the choice of $D_{\rm GMC}=10$ pc, the cell at the $10^{\rm th}$ level has size smaller than $D_{\rm GMC}/3$ at $z>8$. In this case, some fraction of the material in the sphere that lies beyond the 27 cells cannot participate cluster formation. For this reason, we control the refinement criteria so that the smallest cell size is always larger than $\rm D_{GMC}/3$. This suggests that gas cells are not allowed to reach the $10^{\rm th}$ level of refinement until $z<8$.

We also explored the possibility of varying sphere size based on the observed mass-size relation of local GMC \citep[e.g.][]{larson81}. However, it requires several iterations to obtain the appropriate cloud size and sometimes fails to converge. Another uncertainty comes from the normalization of the mass-size relation for GMCs in high-z environment. Therefore, we did not apply it in our current simulations.

As we mentioned above, if there exists an active cluster particle in a cell, the cluster mass will grow with the rate given by Eq.~(\ref{eq:cfr}). The momentum and metallicity increment will also be added to the active particle accordingly. If there are more than one active cluster particle in a cell, we only grow the one with the smallest velocity relative to the host cell. This method prevents adding mass to unrelated cluster particles that occasionally fly through the cell. We also explored an alternative scenario of treating multiple active particles in a single cell by merging them into one active cluster. We found that the CIMF is no longer a power-law shape, but highly inclined toward high masses. This is possibly because that active clusters that fly from other cluster formation regions are absorbed by the main cluster, and clusters in different cluster formation clouds are merged.

Before creating a new cluster particle, we predict its mass accumulated over the timescale ($\tau_{\rm max}$) by assuming a constant SFR: $M_{\rm est} = \tau_{\rm max}\dot{M}$. If this mass is smaller than the threshold mass $M_{\rm th}$, the cluster particle will not be formed in order to avoid small-mass particles. After each global timestep, we remove all inactive cluster particles if their masses are smaller than $M_{\rm th}$, and recycle their mass and momentum back to the gas cells. To justify this process, we compare the scientific results and the computational efficiency with and without particle removal. We found that these inactive low-mass clusters contain only 3-5\% of the stellar mass and have a negligible effect on the overall stellar component of the simulated galaxies, but they contribute $\sim$40\% in paricle number and slow down the code. Therefore, it is appropriate to recycle such particles to improve the computational efficiency.

We would like to understand the accretion history of cluster particles. The typical timestep in our simulations is about 1000 years \footnote{ For a gas cell with size $L_{\rm cell}\approx10$~pc, Courant-Friedrichs-Lewy condition requires a timestep smaller than $\Delta t<L_{\rm cell}/v_{\rm gas}=10^4 {\rm yr} \ (L_{\rm cell}/10 \: {\rm pc}) \ (v_{\rm gas}/1000 \: {\rm km/s})^{-1}$. Due to the discreteness of the timestep refinement among various levels, the timestep at the highest level is even shorter if it is passed from lower level cells with high velocity and high temperature. We find that the typical timestep for our fiducial run is about 1000 yr.}. The active cluster formation period can be resolved by several thousand steps. Since it is not practical to store each local timestep, we record two integral timescales that characterize the mass accretion history. The first is the total duration of a star-forming episode, $\tau_{\rm dur}$, from the time of creation of the cluster particle to the last time before it becomes inactive. The second is the mass-weighted cluster formation timescale:
\begin{equation}\label{eq:timescale}
\tau_{\rm ave} \equiv \frac{\int_{0}^{\tau_{\rm dur}} t \, \dot{M}(t) \, dt}
{\int_{0}^{\tau_{\rm dur}} \dot{M}(t) \, dt},
\end{equation}
where $\dot{M}(t)$ is the mass accumulation rate of a given cluster at time $t$, measured from the moment of particle creation. By construction, the duration $\tau_{\rm dur}$ cannot exceed the maximum formation time $\tau_{\rm max}$. If the mass accumulation history is a power-law, $\dot{M}(t)\propto t^n$, the relationship between $\tau_{\rm ave}$ and $\tau_{\rm max}$ can be estimated as $\tau_{\rm ave} = \tau_{\rm max} \frac{n+1}{n+2}$. For example, a time-independent $\dot{M}$ gives $\tau_{\rm ave} = \tau_{\rm max}/2$. If $\dot{M}$ decreases over time, as a result of self-feedback of newly formed stars, we expect $\tau_{\rm ave}<\tau_{\rm max}/2$. We discuss the accumulation history of star clusters in Section~\ref{sec:timescale}.

\subsection{Dynamical Disruption of Star Clusters}\label{sec:disruption}


Above we described the detailed implementation of the creation and growth of cluster particles. Another key ingredient, especially for the study of globular clusters, is their subsequent dynamical evolution. Star clusters suffer tidal dissolution during the early stage when they are located within the gaseous disk, and experience gradual evaporation of stars after they escape to the galaxy halo. We create a new variable, the bound fraction $f_{\rm bound}$, to represent the fraction of mass that is still bound to the cluster particle at a given time, and update $f_{\rm bound}$ based on the local tidal field as well as the rate of internal evaporation\ \citep{gnedin_etal14}. We assume that the unbound part remains near the cluster, so that the overall stellar mass distribution is unaffected by cluster disruption. Since this paper focuses mainly on the formation process of young star clusters, we will not discuss here the details of calculating cluster evolution, and leave this topic to a follow-up paper.

\subsection{Stellar Feedback}\label{sec:feedback}

\begin{figure*}[t]
\includegraphics[width=1.0\hsize]{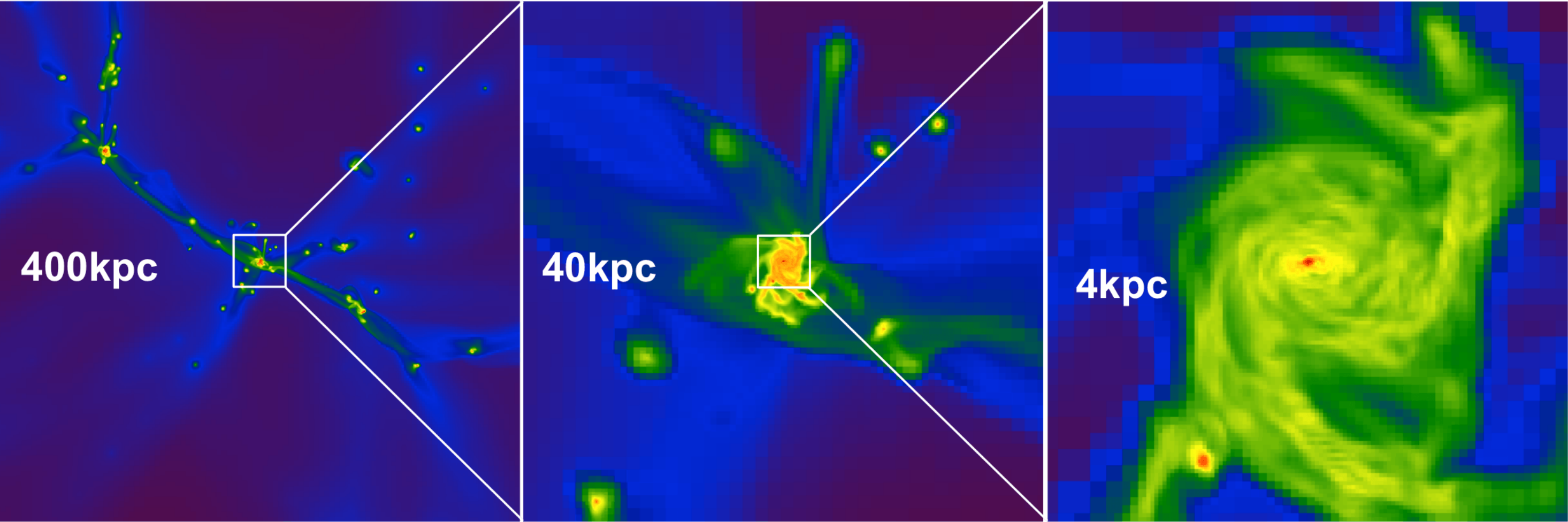}
  \vspace{0cm}
\caption{\small Gas density projection plots of the fiducial run centered on the main galaxy at $z\approx3.3$. The physical sizes of box for the three panels, from left to right, are 400 kpc, 40 kpc, and 4 kpc, respectively.}
  \vspace{0cm}
  \label{fig:projection}
\end{figure*}

The creation of cluster particles is followed by stellar feedback that returns mass, momentum, and energy into the surrounding medium. The feedback process has been demonstrated to be crucial in maintaining steady star formation activity.

The sub-grid stellar feedback model used in our simulations is similar to that described in \citet{agertz_etal13}. It includes the injection of thermal energy, momentum, metals, and ionization radiation from supernova (SN) explosion, stellar winds, stellar luminosity, and radiative pressure onto surrounding gas and dust. The SN rate of a given cluster particle is calculated by assuming the Kroupa initial mass function \citep{kroupa_01}, and the injected energy and momentum are calibrated by Padova stellar evolution models \citep[see details in][]{agertz_etal13}. An analytical fit by \citet{gnedin14} is used to account for the time evolution of the ionization radiation from young stars. During the growth of active cluster particles, we use the mass-weighted age as the age of the stellar population, which determines the amount of feedback. We also include the SGS turbulence model that treats the unresolved turbulent energy as a separate hydrodynamic field. This field is similar to the ``feedback energy variable'' in \citet{agertz_etal13}, but in addition to isotropic non-thermal pressure support it includes anisotropic production of the SGS turbulence by cascade from the resolved scales, turbulent diffusion and dissipation of turbulence over local crossing time rather than fixed rate of decay assumed in  \citet{agertz_etal13} \citep[see details in][]{semenov_etal15, schmidt14}.

The momentum exerted from stellar particles is distributed evenly to 26 nearest neighbors surrounding the parent cell of the stellar particles. The feedback momentum is added directly into the cell if it has the same direction with the momentum of that cell. Otherwise, the feedback momentum is subtracted from the cell momentum and the kinetic energy associated with the canceled momentum is converted to thermal energy and added to the cell internal energy. When adding feedback energy or momentum onto a given cell, we also check whether the temperature (or velocity) of the cell is larger than $T_{\rm fb}^{\rm ceiling}=10^8$~K (or $v_{\rm fb}^{\rm ceiling}=5000$ km s$^{-1}$). To avoid creating unrealistic hot and fast flow, feedback does not add to cells whose temperature or velocity has already reached the ceiling value.

Although the feedback model contains many physical ingredients from various sources, it should be noted that the only adjustable parameter is the fraction of SN energy that is converted into turbulent energy, $f_{\rm turb}$. In the fiducial run, we set $f_{\rm turb}=10\%$ and we vary this factor to 50\% in TURB50 model to test the effect of turbulence on our results.

\section{Global properties of simulated galaxies}\label{sec:result}

The simulations stall at low redshift because of the very short timestep. Our fiducial run has reached $z=3.3$ and most of the results about it are from that snapshot. We use the ROCKSTAR halo finder \citep{behroozi_rockstar} to identify halos and substructures in the whole simulation box, and construct halo merger trees by the consistent-tree algorithm \citep{behroozi_tree}.

Figure \ref{fig:projection} shows the gas density projection plots of the main galaxy for the fiducial run at $z\approx3.3$ in three different scales from 400 kpc to 4 kpc. The filamentary structure that connects galaxies is clearly seen in the left panel. In the central panel, the gaseous disk of the main galaxy is shown in a face-on view, with obvious spiral patterns. There are also substructures orbiting the main galaxy, including significant tidal streams. In the right panel, the simulation is zoomed in to 4 kpc, and the discreteness of outer gas cells becomes visible. In the outer edge of the galaxy, the density is low and the mesh is coarse, while at the center, the highest refinement level is reached due to the high density.

We also find that gas fragmentation due to gravitational instability generates a large number of dense clouds along the spiral arms. Massive star clusters are formed in these clouds. Their radiation ionizes the gas nearby, and some fraction of the radiation escapes the host galaxy and contributes to the extragalactic UV flux that reionized the universe at high redshift \citep[e.g.,][]{gnedin16}. The high spatial resolution and three-dimensional radiative transfer implemented in our simulations make it possible for us to study the escape of radiation from young clusters, and in particular, estimate the escape fraction in the inner few hundred parsecs of the galaxy. Since the main goal of this paper is to study the properties of star clusters, we present a preliminary analysis of the escape fraction in Appendix.

\begin{figure}[t]
\includegraphics[width=1.0\hsize]{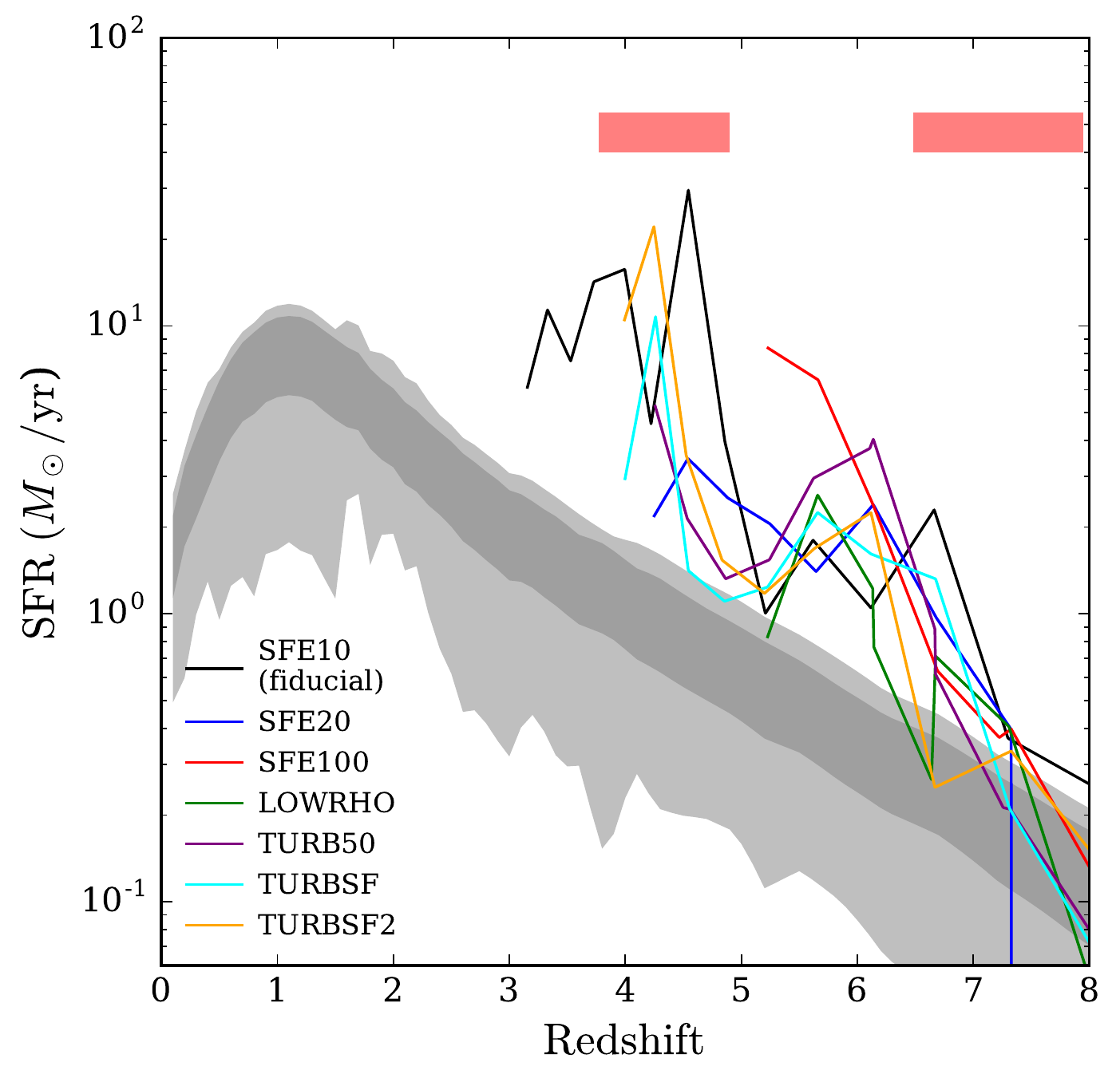}
  \vspace{-0.3cm}
\caption{\small Simulated star formation history of the main galaxy, smoothed over 100 Myr bins, for different models (see legend). The star formation history for a $M_{\rm vir}(z=0) = 10^{12}M_\sun$ halo from the abundance matching technique \citep{behroozi_etal13} is overplotted by shaded regions. The dark and light regions show the one- and two-sigma confidence intervals, respectively. Red horizontal bars show the epochs of two major-merger events (with mass ratio larger than 0.3) that are identified from the merger trees of the main halo. The span of the bar represents the duration of each merger, see Sec. \ref{sec:merger} for detailed description.}
  \vspace{0cm}
  \label{fig:sfh}
\end{figure}

Figure~\ref{fig:sfh} shows the star formation histories (SFH) of the main progenitor for different models, compared to the SFH of an average $M_{\rm vir}(z=0) = 10^{12}\, M_\sun$ halo derived from the abundance matching technique \citep{behroozi_etal13}. The star formation rate is calculated from the cluster particles within the main galaxy at each snapshot and smoothed over 50~Myr. 

A similar feature of all SFHs is the general rise towards lower redshift, with periodic bursts of star formation. Two noticeable bursts at $z \approx 6-7$ and $z \approx 4-5$ coincide with two major-merger events that are identified from the merger trees. The star formation rates reflect a particular mass assembly history encoded in the initial conditions of our simulations, but also depend on the model parameters such as $\epsilon_{\rm ff}$ and $f_{\rm turb}$. For example, the peaks of the bursts shift to later times with decreasing $\epsilon_{\rm ff}$ and increasing $f_{\rm turb}$, and the fiducial run shows significant reduction of SFR after the second burst.

It also appears that the feedback in our simulations is not strong enough to match the average SFR predicted by \citet{behroozi_etal13}. It could be due to the implementation of feedback processes, or the specific initial condition we chose for these runs. As we described in Section~\ref{sec:feedback}, subgrid turbulence in our simulations decays over the local crossing time. For strongly turbulent cells, this timescale can be much shorter than 10 Myr, a constant dissipation time that was assumed in \citet{agertz_kravtsov_15}. As a result, the sub-grid turbulence decays faster in our model and the turbulent pressure becomes insufficient to push the gas away from the cluster-forming regions. Therefore, the dynamical effect of the sub-grid turbulence in our simulations is not as significant as that in \citet{agertz_kravtsov_15}. In addition, even simulations with stronger feedback implementation, such as FIRE \citep{hopkins_etal14}, show a larger range of SFH variation than predicted by the abundance matching. For example, their run m12q has higher SFR at high redshift due to a particular choice of "quiescent" initial conditions.

\section{Cluster initial mass function}\label{sec:CIMF}

\subsection{Power law vs. Schechter function}\label{sec:schechter}

One of the most fundamental properties of young star clusters is the CIMF. Observations of nearby galaxies have found that CIMF follows the Schechter function form: a power-law distribution with an exponential cutoff at $M_{\rm cut}$ \citep[see][and references therein]{portegies_zwart_etal10}:
\begin{equation}\label{eq:cimf}
\frac{dN}{dM}\propto M^{-\alpha}\exp{(-M/M_{\rm cut})}.
\end{equation}
The power-law index lies in a fairly narrow range $\alpha \approx 1.8-2.2$, gradually getting steeper for less massive galaxies. The cutoff mass varies significantly more from galaxy to galaxy and scales most strongly with the star formation rate of the host galaxy. The typical value of $M_{\rm cut}$ for Milky Way-sized spiral galaxies is about $2\times10^5\, M_\odot$ \citep[e.g.][]{gieles_etal06}, increasing to $M_{\rm cut}>10^6\, M_\odot$ for luminous interacting galaxies \citep{bastian_08}. 

\begin{figure}[t]
\includegraphics[width=1.0\hsize]{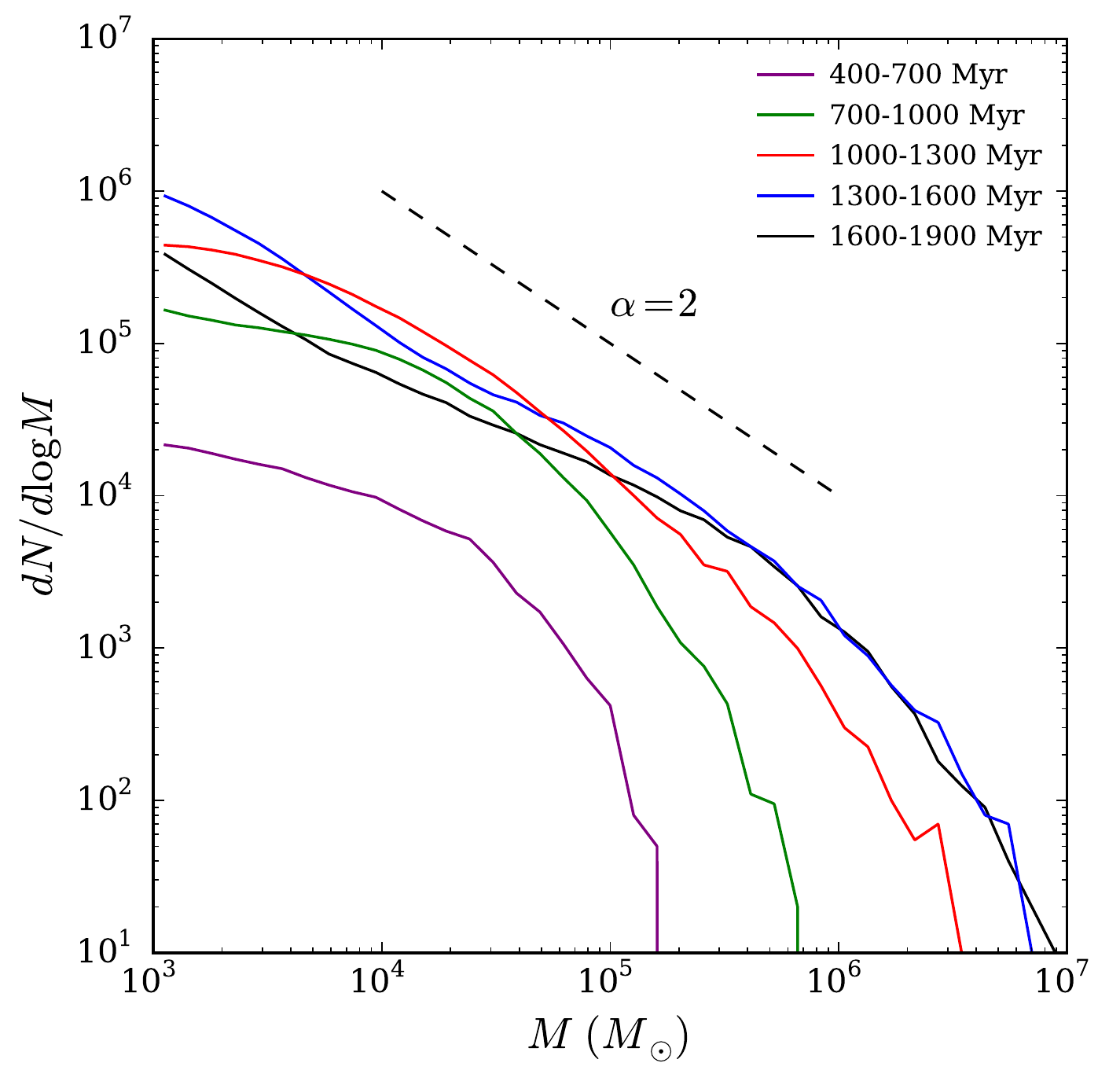}
  \vspace{-0.3cm}
\caption{\small Cluster initial mass function for different cluster age bins in the main galaxy for the fiducial run. Cluster disruption is not included. A power-law distribution with slope of 2 is plotted by a dashed line. The mass functions exhibit a stable shape across a large range of cosmic time, and both the power-law slope and the high mass cutoff are consistent with observed cluster samples in nearby galaxies.}
  \vspace{0cm}
  \label{fig:CIMF_fiducial}
\end{figure}

Figure~\ref{fig:CIMF_fiducial} shows the model CIMF in the main galaxy for the fiducial run. The CIMF exhibits the Schechter function-like shape with a power-law slope similar to the observed. The normalization and cutoff mass vary with cosmic time, increasing at least until $z=4$.  Both variations are correlated with the galaxy SFR.  All of these trends are consistent with the observations of young star clusters, which means our implementation of cluster formation, at least in its general features, is realistic.

\begin{figure}[t]
\includegraphics[width=1.0\hsize]{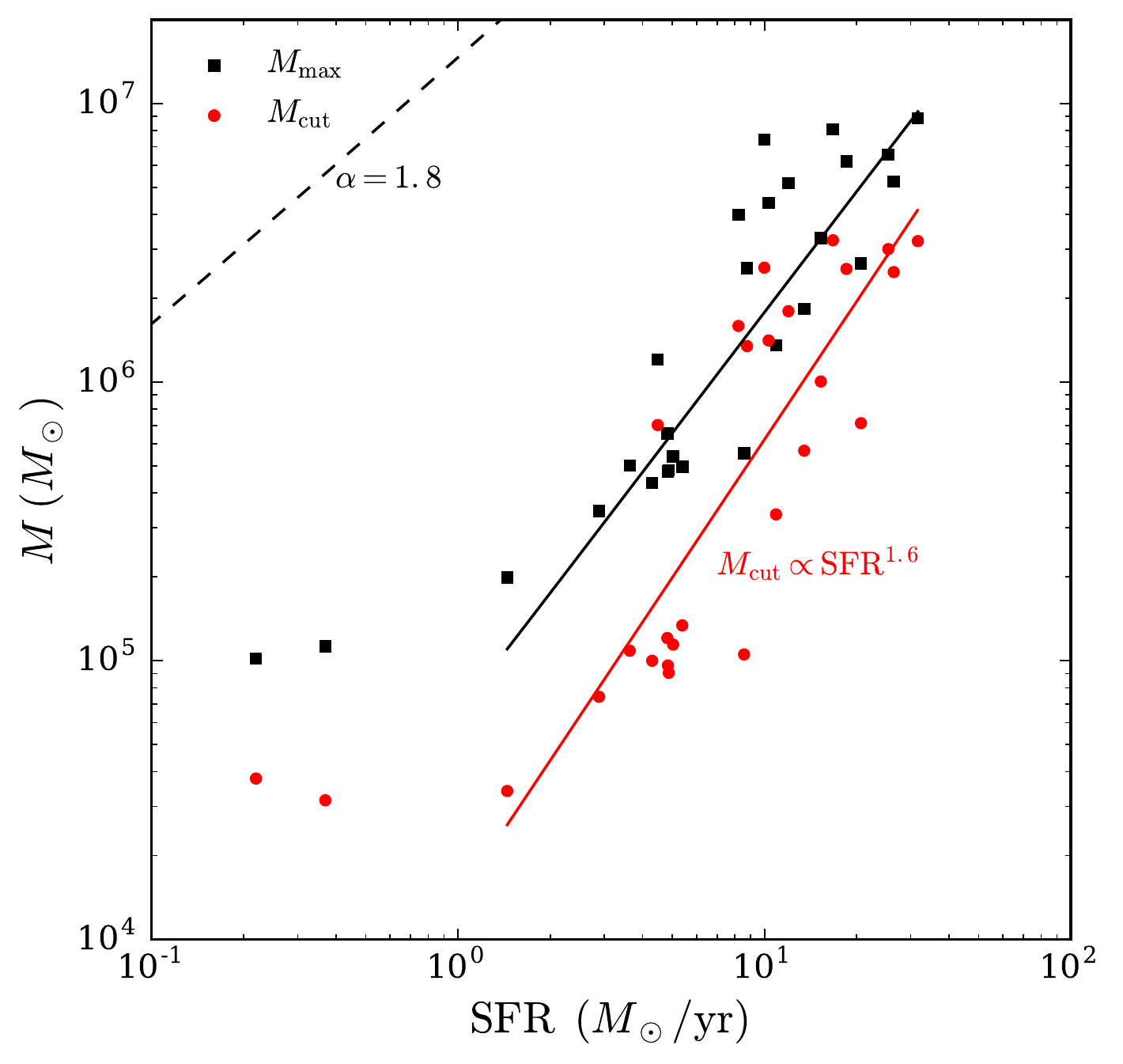}
  \vspace{-0.2cm}
\caption{\small SFR vs. maximum cluster mass ($M_{\rm max}$) and cutoff mass ($M_{\rm cut}$) for the fiducial run. The SFR is averaged over 50 Myr and $M_{\rm max}$ is chosen from the clusters found in the same time interval. The initial mass function of clusters in each interval is fitted by Eq. (\ref{eq:cimf}), and both the power-law slope and cutoff mass are obtained. The red solid line shows the best-fit relation between $M_{\rm cut}$ and SFR for samples with ${\rm SFR} > 1 M_\odot\, {\rm yr}^{-1}$. The theoretical maximum masses for given SFRs for both pure power-law and Schechter mass function are shown by dashed and solid black lines, respectively (See Section \ref{sec:schechter} for detailed calculations). Note that the black solid line is not the best-fit between SFR and $M_{\rm max}$, but the expected $M_{\rm max}$ by assuming the Schechter CIMF with $\alpha=1.8$ and $M_{\rm cut}$ that is estimated by the empirical $M_{\rm cut}$-SFR relation (red line).}
  \vspace{0.2cm}
  \label{fig:sfr_mass}
\end{figure}

\begin{figure}[t]
\includegraphics[width=1.0\hsize]{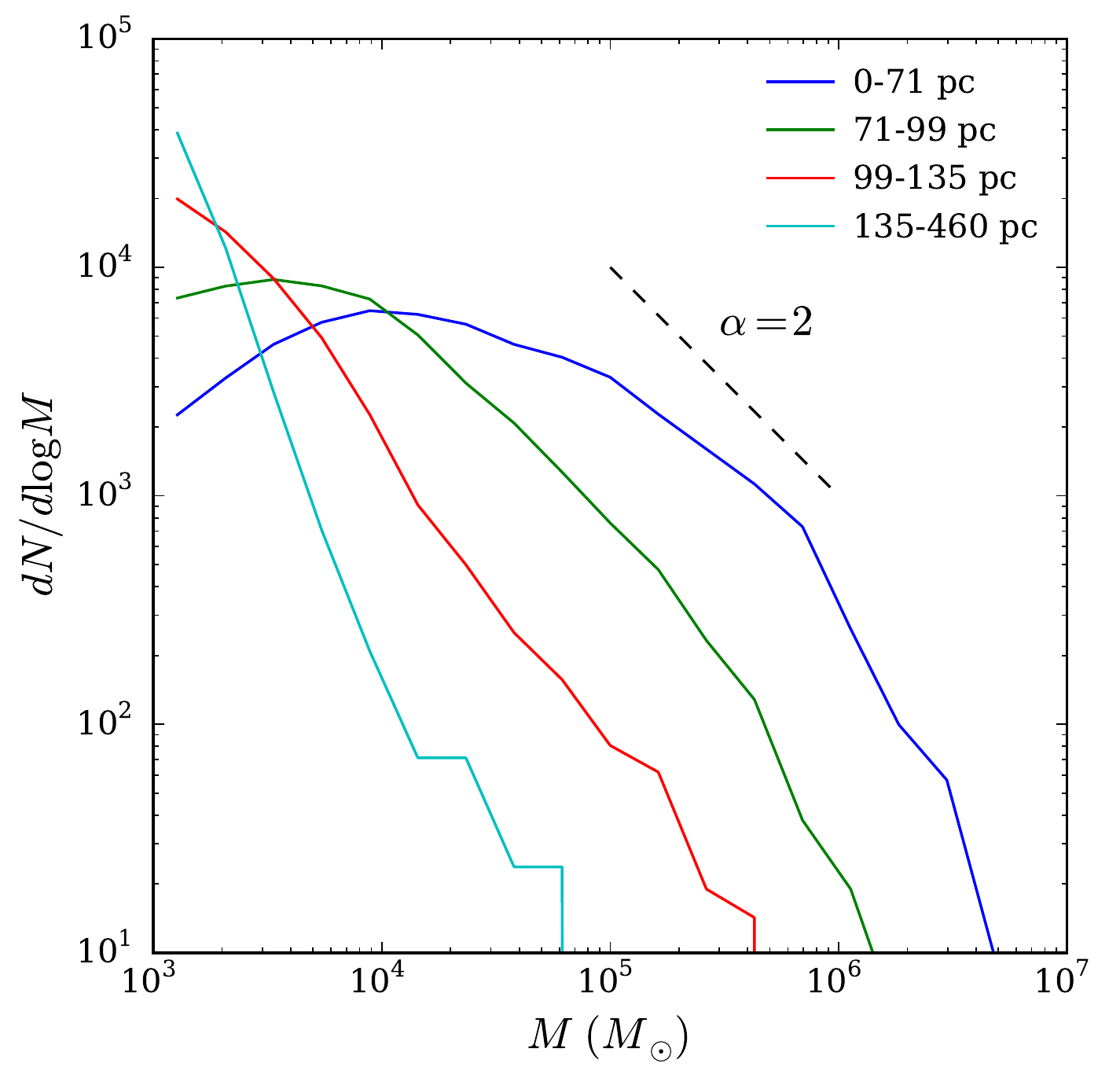}
  \vspace{-0.2cm}
\caption{\small Initial mass function of young clusters (<100 Myr) in four radial bins of equal cluster number, for the main galaxy in the fiducial run. The mass functions show clear steepening from the inner annulus (blue line) to the outer ones as well as the decreasing maximum cluster mass with radius.}
  \vspace{0.2cm}
  \label{fig:spatial}
\end{figure}

To investigate the relation between SFR and the high mass end of the CIMF at different epochs, we make bins of clusters formed within 50 Myr of each other.  As this interval is more than three times the longest cluster formation time, we can treat each bin as a roughly independent measurement.  For each bin, we estimate the average SFR, maximum cluster mass $M_{\rm max}$, the best-fit power-law slope $\alpha$, and cutoff mass $M_{\rm cut}$. The average value of the best-fit slope for all bins is $\alpha\approx 1.8$, similar to the observed.

Figure~\ref{fig:sfr_mass} shows a strong correlation between the SFR and $M_{\rm max}$, as well as $M_{\rm cut}$, for bins with ${\rm SFR} > 1\, M_{\odot}/{\rm yr}$. The best-fit power law relationship between SFR and $M_{\rm cut}$ is:
\begin{equation}\label{eq:Mcut_SFR}
M_{\rm cut} \approx 1.4\times10^4\, M_\odot \left(\frac{\rm SFR}{1M_\odot/{\rm yr}}\right)^{1.6},
\end{equation}
and the relationship between SFR and $M_{\rm max}$ is:
\begin{equation}\label{eq:Mmax_SFR}
M_{\rm max} \approx 8.8\times10^4\, M_\odot \left(\frac{\rm SFR}{1M_\odot/{\rm yr}}\right)^{1.4}.
\end{equation}
At ${\rm SFR} < 1\, M_{\odot}/{\rm yr}$, the small samples of clusters may hinder accurate determination of the mass function shape.

Theoretically, for samples drawn from a given mass function, the maximum mass can be calculated by the integral equation: $N(\ge M_{\rm max}) = 1$. Comparing the estimated $M_{\rm max}$ with the actual value can help us validate or rule out specific functional forms of $dN/dM$. The normalization of CIMF comes from the total mass of clusters formed within our chosen time interval, which is simply $M_{\rm tot} = \mathrm{SFR} \times 50$~Myr. Then it is only the shape of CIMF that determines the maximum cluster mass.

For the case of a pure power-law, $dN/dM \propto M^{-\alpha}$, with $1 < \alpha < 2$ and $M_{\rm min} \ll M_{\rm max}$, simple integration gives $M_{\rm max} \approx \frac{2-\alpha}{\alpha-1}\, M_{\rm tot}$. From Figure~\ref{fig:sfr_mass}, we can see that this theoretical $M_{\rm max}$ overestimates the actual maximum cluster mass in the simulations by more than one order of magnitude.

Alternatively, for the case of the Schechter function CIMF as described by Eq.~(\ref{eq:cimf}), the relationship between $M_{\rm tot}$ and $M_{\rm max}$ is
\begin{equation}\label{eq:Mmax_ple}
M_{\rm tot} =  \frac{\Gamma(2-\alpha, M_{\rm min}/M_{\rm cut})-\Gamma(2-\alpha, M_{\rm max}/M_{\rm cut})}{\Gamma(1-\alpha, M_{\rm max}/M_{\rm cut})} M_{\rm cut} ,
\end{equation}
where $\Gamma(s, x)$ is the upper incomplete Gamma function.
Taking $\alpha=1.8$, $M_{\rm min}=10^3\, M_\odot$, and the mean relation between SFR and $M_{\rm cut}$ from Eq. (\ref{eq:Mcut_SFR}), the expected $M_{\rm max}$ can be obtained by solving Eq. (\ref{eq:Mmax_ple}) numerically. The result is shown by solid black line in Figure~\ref{fig:sfr_mass}.  An approximate solution is $M_{\rm max} \approx M_{\rm cut}\, \ln{(M_{\rm tot}/M_{\rm cut}/\Gamma(2-\alpha))}$. This value is much less sensitive to the total mass than the power-law CIMF, and is closer to the cutoff mass.

It should be emphasized that the black line in Figure~\ref{fig:sfr_mass} is not the best-fit relation between SFR and $M_{\rm max}$, but the expected $M_{\rm max}$ calculated from the above procedure. The similarity between this expected $M_{\rm max}$ and the one obtained from the simulations suggests that the Schechter function is an excellent representation of the CIMF in our simulation, while a pure power-law distribution is ruled out. 

\subsection{Spatial variation}\label{sec:spatial}

Another important aspect of CIMF is its spatial variation. Recently, \citet{adamo_etal15} examined the mass function of young star clusters in M83 using multiband \textit{Hubble Space Telescope} imaging data. They split the whole cluster sample into four radial bins and found significant steepening of the CIMF in the outer bins, as well as decrease in the maximum cluster mass. Both trends were related to the steady decrease of the surface density of SFR with radius.

Following a similar procedure, we divide our main galaxy at $z\approx3.3$ by four concentric circles so that each annulus contains the same number of star clusters with age younger than 100 Myr. The CIMF of each bin is shown in Figure~\ref{fig:spatial}. The mass function no longer has a universal shape, but presents a spectrum of power-laws with various slope: the inner bins ($R<100 \, \rm pc$) have $\alpha<2$, while the outer bins ($R>100 \, \rm pc$) have $\alpha>2$, in the mass range $10^4-10^6\, M_{\odot}$. The trend of the CIMF steepening with galactocentric distance is consistent with the spatial variation of CIMF in M83. Moreover, the maximum mass of clusters in the inner bins is much higher than that in the outer ones. The SFR density for the four bins from inner to outer ones is 460, 14, 0.06, and 0.02 $M_{\odot}\, {\rm yr}^{-1}\, {\rm kpc}^{-2}$, respectively. It is clear that the gradient of CIMF correlates with the local SFR density. It is a local manifestation of the same global trend shown in Figure~\ref{fig:sfr_mass}.

\begin{figure}[t]
\includegraphics[width=1.0\hsize]{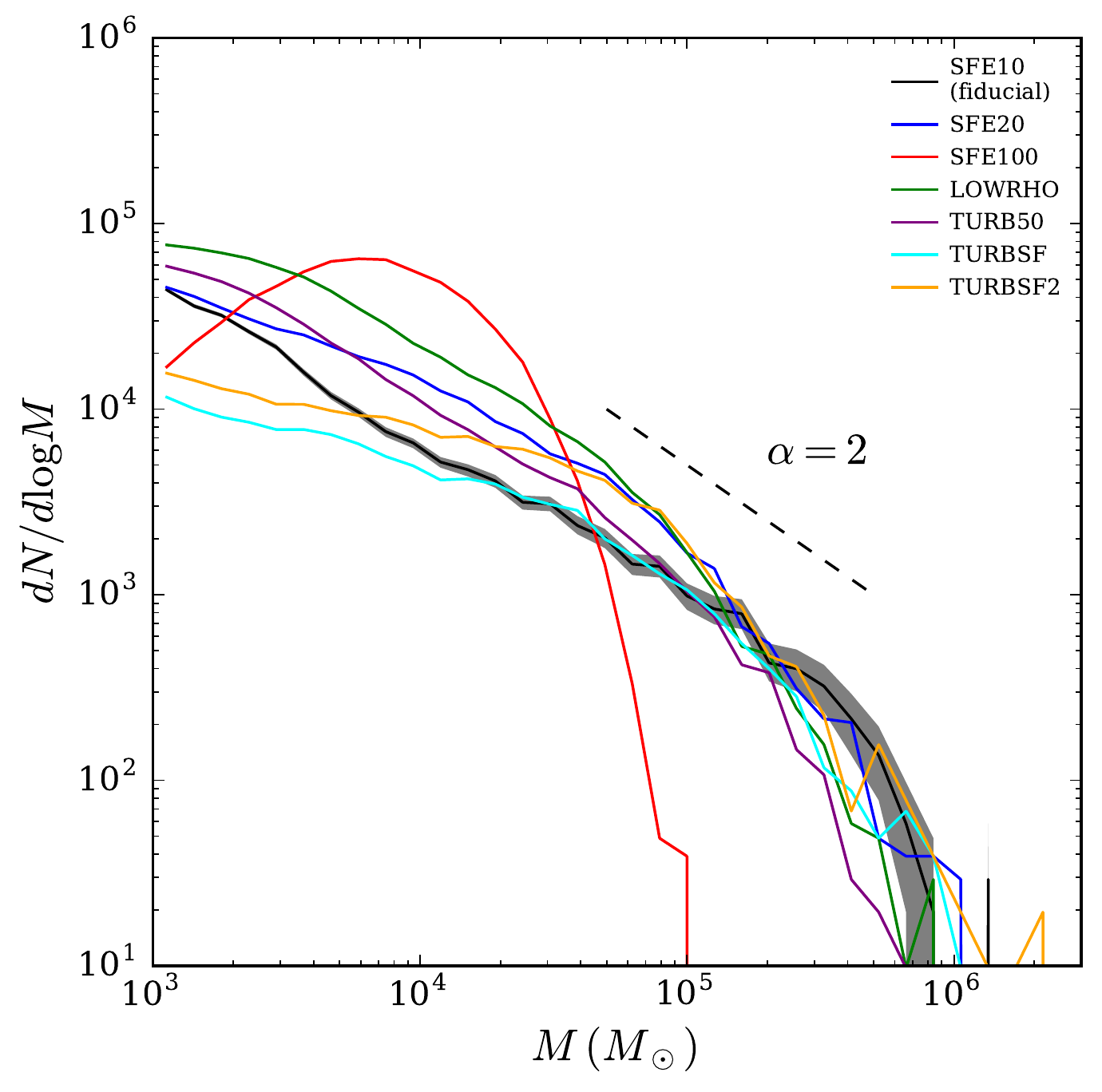}
  \vspace{-0.2cm}
\caption{\small Cluster initial mass function of all seven models (see legend for color codes) in the main galaxy at the same epoch ($z\approx5.3$). The grey shaded region around the black line is the 95\% confidence interval of the CIMF constructed by bootstrap resampling for the SFE10 model (fiducial).}
  \vspace{0.2cm}
  \label{fig:CIMF_models}
\end{figure}

\subsection{Dependence on models of star formation and feedback}

We now explore the variation of CIMF under different model parameters. In Figure~\ref{fig:CIMF_models}, we show the CIMF of all models in the main galaxy at the same epoch. It is a higher redshift ($z \approx 5.3$) than previous plots for the fiducial model ($z \approx 3.3$) because the other runs did not advance as far in time, for reasons of computational efficiency.

Models SFE20 and TURBSF present similar CIMF to the fiducial run, indicating that our cluster formation prescription is not sensitive to model parameters when they are in a reasonable range. In detail, the slope at $M < 10^5\, M_{\odot}$ steepens somewhat with increasing $\epsilon_{\rm ff}$ and, for the one model we explored (TURB50), with increasing $f_{\rm turb}$. In all these cases the range of $\alpha$ is consistent with observations of different nearby galaxies. At high cluster masses the models converge even closer.

The only exception is Model SFE100, with 100\% star formation efficiency per free-fall time, which has a dramatically different CIMF. It deviates from a power-law distribution at low mass, and has the truncation mass that is much lower than the other runs. Although the inability to create massive clusters with a very high star formation efficiency seems counter-intuitive, it reflects the complex nature of the interplay between star formation and feedback: very high star formation efficiency leads to an early starburst; stellar feedback from this burst is strong enough to destroy molecular gas and expel material from the star-forming region; in a short time, the cluster growth is terminated and the final cluster mass is determined only by the first starburst episode. This scenario is confirmed by the analysis of cluster formation timescales in Section~\ref{sec:timescale}.

The differences between the low-SFE (TURBSF, TURBSF2) and high-SFE runs (SFE10, SFE20, TURB50) are reinforced at later times. These five runs have the last common output epoch at $z\approx 4.3$.  The high-SFE runs have a slightly steeper slope $\alpha > 2$ at $M > 10^4\, M_\odot$, which gradually turns over at lower cluster mass and reaches $\alpha = 2$ at $M \approx 10^3\, M_\odot$.  The low-SFE runs show the turnover of the slope to $\alpha < 2$ already at $M \sim 10^5\, M_\odot$.  Lower SFE and weaker feedback runs also reach larger maximum cluster mass, up to a factor of two.  These differences are consistent with the overall trend that more extended star formation events lead to larger eventual cluster mass.

\begin{figure}[t]
\includegraphics[width=1.0\hsize]{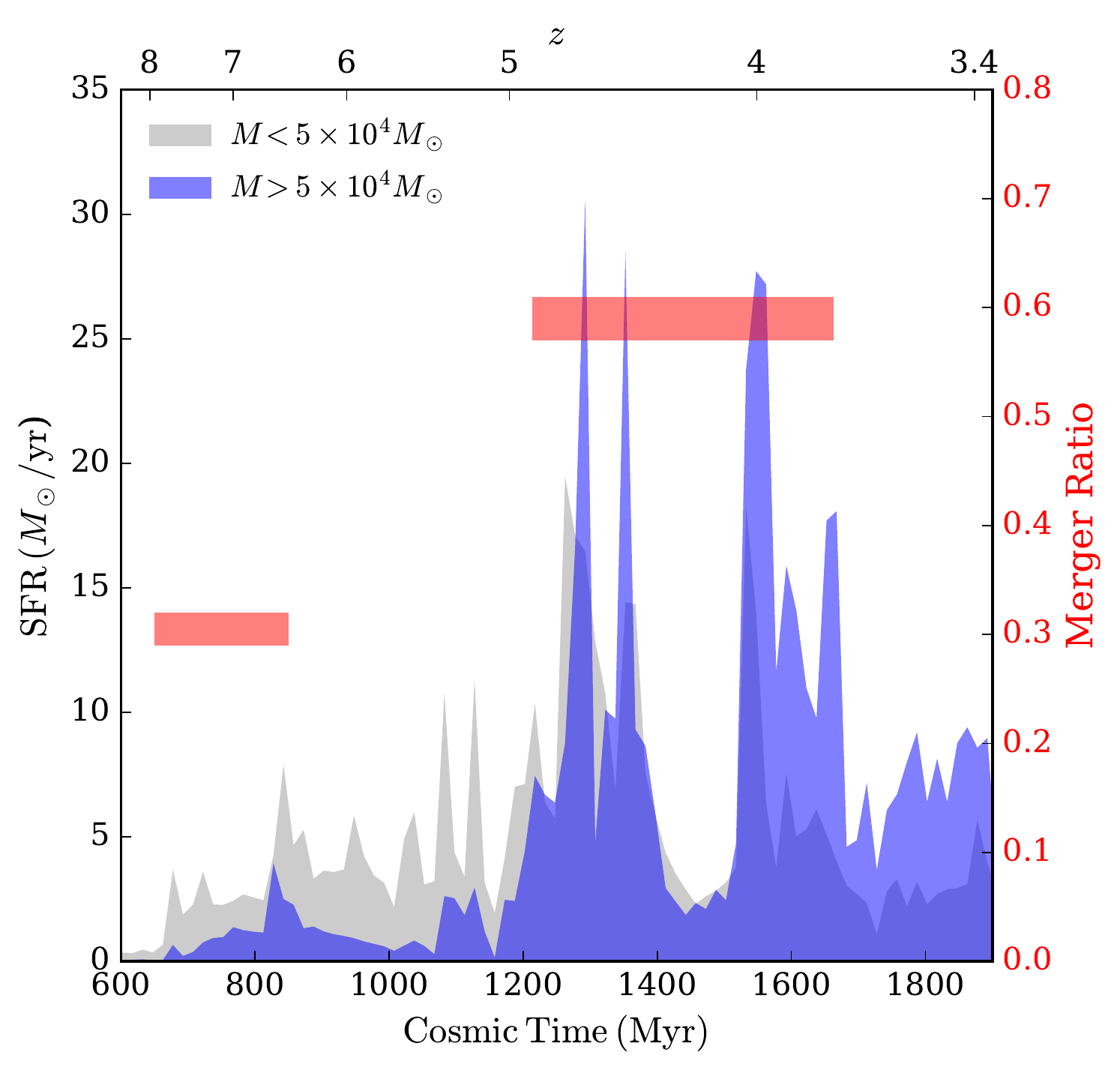}
  \vspace{-0.2cm}
\caption{\small Star formation history of the main galaxy split into massive (blue shaded; $M>5\times10^4M_\sun$) and less massive clusters  (gray shaded; $M<5\times10^4M_\sun$) for the fiducial run. Instead of calculating SFR of the main galaxy at each snapshot as in Section~\ref{sec:result},  here we calculate the formation history of all cluster particles located within the main galaxy at the last snapshot. The SFR is smoothed over 15 Myr. Two major merger events with mass ratio larger than 0.3 are labeled by red bars. The duration of merger is represented by the horizontal length of the bar, while the merger mass ratio that is indicated by the vertical position of the bar according to the scale on the right y-axis.}
  \vspace{0.2cm}
  \label{fig:merger}
\end{figure}

\subsection{Dependence on galaxy mergers}\label{sec:merger}

To investigate the environmental dependence of CIMF, we plot in Figure \ref{fig:merger} the relationship between the star formation history of the main galaxy and the major merger events of its host halo. The SFR is split in two parts, contributed by low-mass and high-mass clusters. The merger ratio is defined as the differential increase of the main halo mass between adjacent snapshots: $R_m \equiv (M_{h,i}-M_{h,i-1})/M_{h,i-1}$, where $M_{h,i}$ is the mass of the main halo in the $i^{\rm th}$ snapshot. The progenitor and descendant information of the main halo is extracted from the merger tree calculated by ROCKSTAR. Only mergers with $R_m>0.3$ are shown in the Figure, and the durations of the two merger events are determined by visual inspection of the dark matter density distribution across several snapshots. Quantitatively, we set the starting point of a merger as the time when the host-satellite separation becomes less than $\sim 30$~kpc, following \citet{behroozi_etal15} who found systematic enhancement of SFR for galaxy pairs with projected separation smaller than about 60~kpc, but not at larger separations. 

We can verify indeed that the last merger with mass ratio $R_m=0.6$ at $z\approx4-5$ triggers starbursts with SFR peaks around $30\, M_\sun\, {\rm yr}^{-1}$, and creates a large fraction of massive clusters. There are several distinct bursts, at least 5, in the time interval of about 400~Myr that it takes to complete the merger. They are likely associated with two orbital passages of the merging galaxies and gravitational instabilities in the new combined gaseous disk. The first merger at early times ($z\approx 7$) does not produce many massive clusters, as the galaxies are still too small to contain sufficient amount of cold gas. Some massive clusters also form in the quiescent period following the last merger, when the galactic disk still holds large gas reservoir, and so the merger activity and associated potential perturbations are not unique requirements to create massive clusters. Still, our results show that the rate of cluster formation is enhanced during the periods of gas-rich major mergers.

\begin{figure}[t]
\includegraphics[width=1.0\hsize]{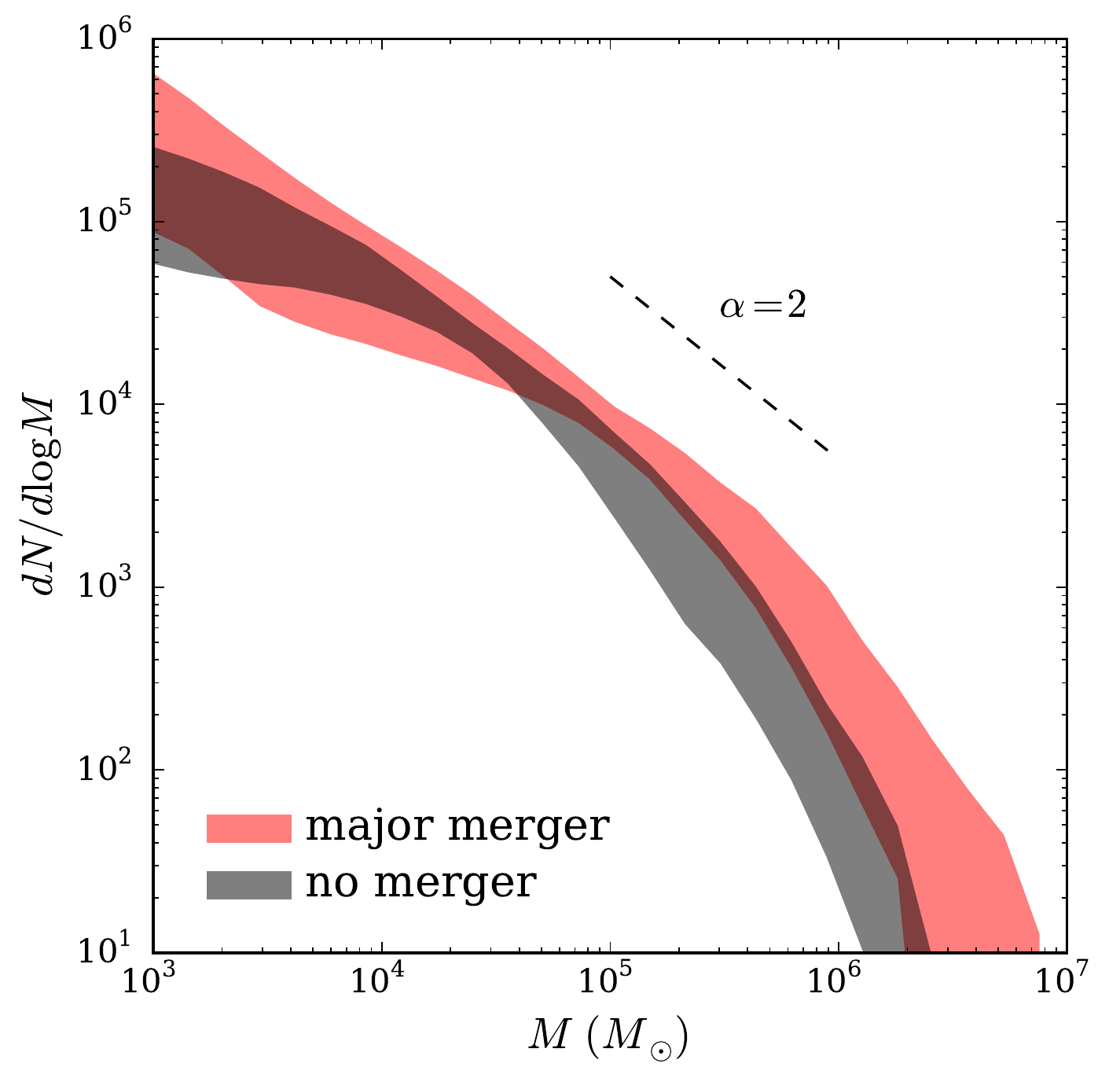}
  \vspace{-0.3cm}
\caption{\small Cluster initial mass function split by merger activity: the mass function during major merger (black) and between mergers (red). Each band shows the standard deviation of the mass functions around the mean value of models SFE10 SFE20, TURB50, TURBSF, and TURBSF2.}
  \vspace{0.2cm}
  \label{fig:CIMF_merger}
\end{figure}

To investigate possible differences in the shape of CIMF during major mergers versus quiescent periods, we split the model clusters into merger-generated ($t_{\rm creation}=1250-1400$~Myr) vs. non-merger-generated ($t_{\rm creation}=900-1200$~Myr) groups. Figure~\ref{fig:CIMF_merger} shows that the cluster mass distribution has higher cutoff mass during the major merger event. This can be understood using the empirical ${\rm SFR}-M_{\rm cut}$ relation described by Eq.~(\ref{eq:Mcut_SFR}): $M_{\rm cut} \propto {\rm SFR}^{1.6}$. In the non-merger case, the typical SFR is about $5\, M_\odot\, {\rm yr}^{-1}$ and the corresponding cutoff mass is $1.8\times10^5\, M_\odot$, whereas in the merger case, $M_{\rm cut}=1.7\times10^6\, M_\odot$ for ${\rm SFR}=20\, M_\odot\, {\rm yr}^{-1}$. The ratio of the two cutoff masses is almost exactly $(20/5)^{1.6}$. Physically, this trend is consistent with the picture of the formation of massive star clusters in high-pressure environment that is produced by major mergers \citep[e.g.,][]{muratov_gnedin10, kruijssen14, renaud_etal15}.

\subsection{Dependence on density of star formation rate}

Analytical models of cluster formation predict that the fraction of clustered star formation increases with the intensity of star formation; specifically, with the SFR per unit area, $\Sigma_{\rm SFR}$, not just the total SFR in a given galaxy.  In order to investigate the relationship between SFR density and the cluster fraction, $\Gamma$, we need to define $\Gamma$ in the simulation analogously to how it is measured in observations.  This is hard to do, because we represent all stellar distribution as clusters of various mass, whereas in observations clusters are typically determined using group-finding algorithms applied to continuous stellar maps.  As a proxy for $\Gamma$ we choose the fraction of all star formation contained in clusters more massive than a given mass $M_{\rm cl}$:
\begin{equation}
\Gamma_{\rm sim} \equiv \frac{\rm SFR(M>M_{\rm cl})}{\rm SFR}.
\end{equation}
It is not the same quantity as the observed $\Gamma$, and therefore, comparison between the two should be treated with caution. Our goal is to study the trend of $\Gamma_{\rm sim}$ with $\Sigma_{\rm SFR}$, and only loosely related it to the observed range of $\Gamma$.

We split the main and satellite galaxies into concentric circular bins, as we did to study the spatial variation of CIMF in Section~\ref{sec:spatial}. We calculate $\Sigma_{\rm SFR}$ by dividing the SFR by the area of the annulus. We calculate the SFR by counting all cluster particles in a given annulus with ages between 15 to 50 Myr and use $M_{\rm cl} = 10^4\, M_\odot$.  This is a compromise value for the different choices used in the studies of star clusters in M83 (roughly $M > 10^3\, M_\odot$) by \citet{adamo_etal15} and seven other nearby galaxies (roughly $M > 10^{4.7}\, M_\odot$) by \citet{goddard_etal10}. Including the cluster disruption would not affect our calculation, as the mass loss over only 50 Myr is expected to be minimal for all clusters more massive than $10^3\, M_\odot$.

\begin{figure}[t]
  \vspace{0cm}
\includegraphics[width=1.0\hsize]{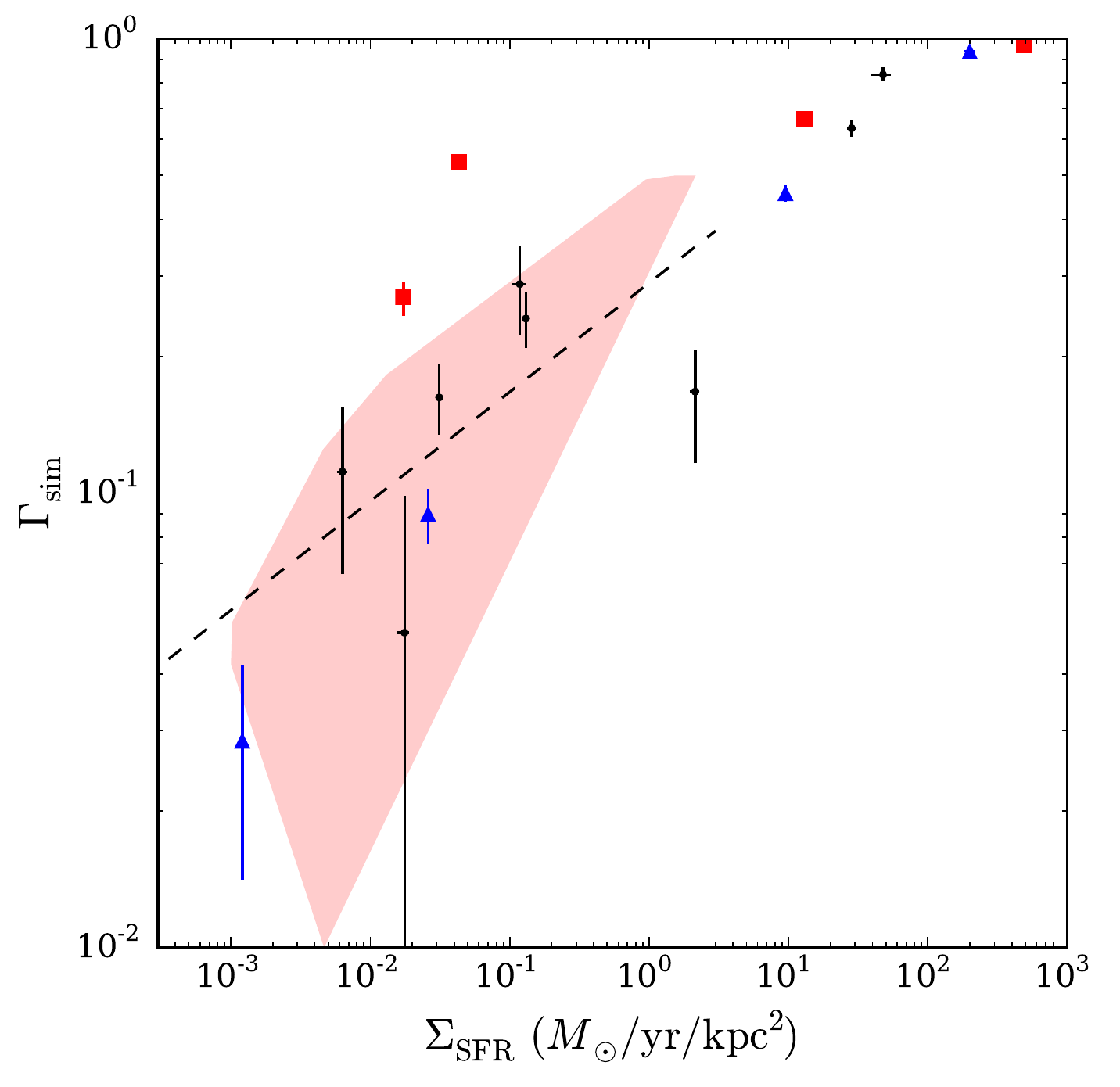}
  \vspace{-0.3cm}
\caption{\small SFR density vs. fraction of young massive star clusters ($M > 10^4\, M_\odot$) in both main and satellite galaxies for the fiducial run at $z\approx3.3$. The red and blue points represent the clusters in the main and the second largest galaxies, respectively. Black points show the other satellite galaxies. Dashed line shows the empirical relation for the observed star cluster populations in 7 nearby galaxies by \citet{goddard_etal10}, while the pink shaded region shows the envelope that covers the data points compiled in \citet{adamo_etal15}.}
  \vspace{0.2cm}
  \label{fig:sfr_cfe}
\end{figure}

Figure~\ref{fig:sfr_cfe} shows the relationship between the SFR surface density and $\Gamma_{\rm sim}$ for the fiducial run at $z \approx 3.3$. The trend that higher $\Sigma_{\rm SFR}$ leads to higher $\Gamma_{\rm sim}$ is reproduced in our model over a large range of $\Sigma_{\rm SFR}$ from $10^{-3}$ to $10^{3}\, M_\sun\, {\rm yr}^{-1}\, {\rm kpc}^{-2}$. To test the robustness of this result, we vary $M_{\rm cl}$ from $5\times10^3\, M_\odot$ to $10^5\, M_\odot$ and find that, although the absolute value of $\Gamma_{\rm sim}$ changes with $M_{\rm cl}$, the positive correlation remains. We also find that, for the same galaxy, both $\Sigma_{\rm SFR}$ and $\Gamma_{\rm sim}$ are higher in the inner annuli than in the outer ones. This is consistent with the spatial variation of CIMF, as we discussed in Section~\ref{sec:spatial}. The shallower slope and higher cutoff mass of CIMF in the inner annuli naturally lead to higher cluster fraction.

\section{Cluster formation timescale}\label{sec:timescale}

\begin{figure*}[t]
  \vspace{0cm}
\includegraphics[width=1.0\hsize]{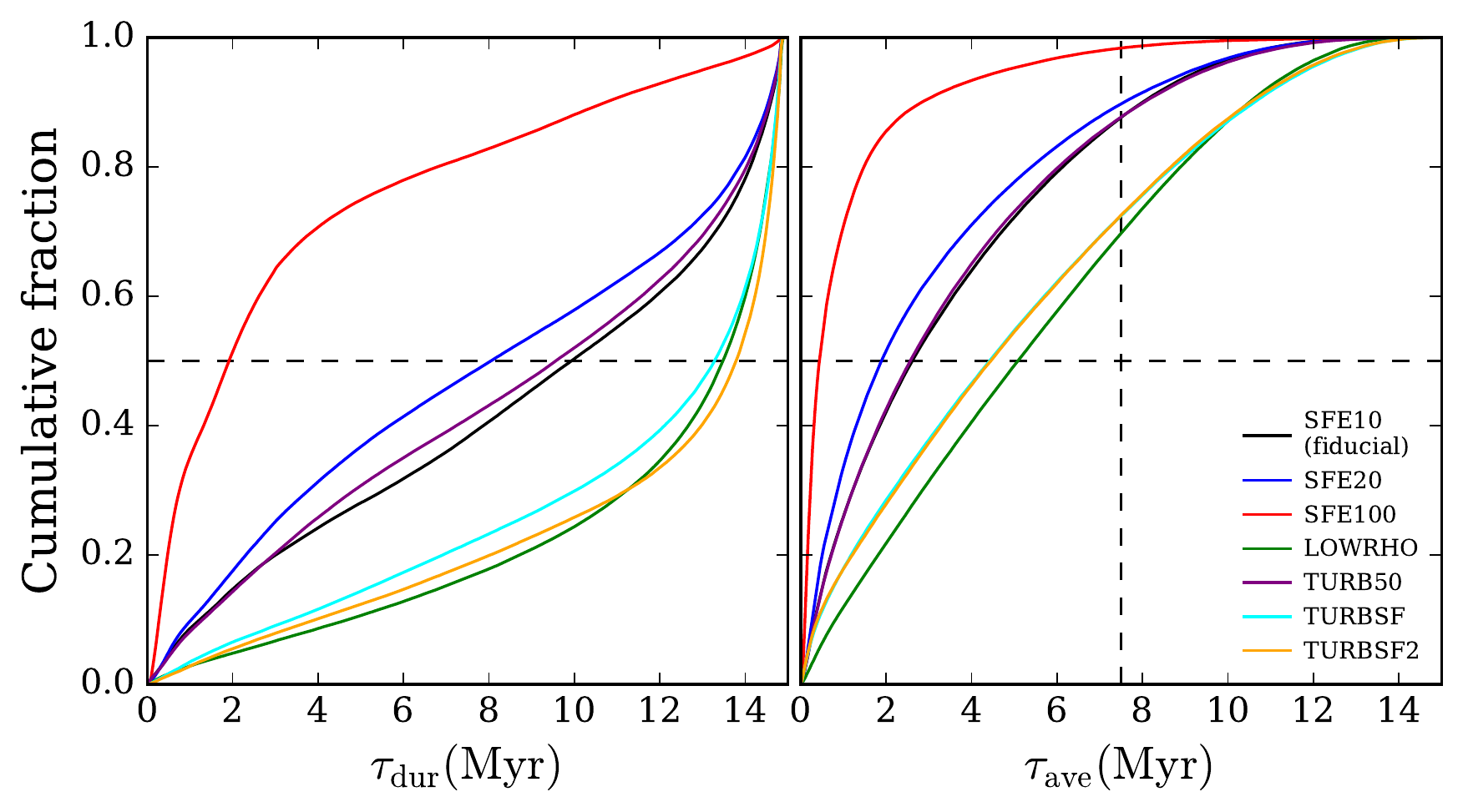}
  \vspace{-0.3cm}
\caption{\small Cumulative distribution function of the cluster formation duration $\tau_{\rm dur}$ (left panel) and the mass-averaged cluster formation timescale $\tau_{\rm ave}$ (right panel). Different colors correspond to the different models described in legend. The cluster samples are selected within the main galaxy at $z\approx5.3$ for all models.}
  \vspace{0.2cm}
  \label{fig:cluster_time}
\end{figure*}

Observations of star-forming regions in the Galaxy suggest that cluster formation process is quick \citep{lada_lada03, maclow_klessen_04}. \citet{hartmann_etal12} compile age information of young stars in the Orion Nebula Cluster and find the age spread as short as only a few Myr. Such a narrow spread presents an additional test of the implementation of cluster formation and feedback.

We investigate the formation timescale and mass accretion history of model clusters by analyzing the distribution of two characteristic timescales, $\tau_{\rm dur}$ and $\tau_{\rm ave}$, defined by Eq.~(\ref{eq:timescale}) in Section~\ref{sec:CCF}. The first is the full duration of the cluster formation episode, from the first star to the last. The second timescale is weighted by star formation rate and more closely corresponds to the observed age spread.

Figure~\ref{fig:cluster_time} shows the cumulative probability distribution of  $\tau_{\rm dur}$ and $\tau_{\rm ave}$ of all cluster particles in the main galaxy for all models at the last common epoch, $z\approx5.3$. As can be seen from the left panel, more than half of the cluster particles stop accretion completely within 10 Myr for models SFE10, SFE20, SFE100, and TURB50. For model SFE100, majority of the cluster particles become inactive within only 2 Myr.  This is because feedback from active cluster particles heats up and removes the gas from their natal GMCs and changes the cell properties so that one or more star formation criteria are violated. For two turbulence-based star formation efficiency models, TURBSF and TURBSF2, star clusters are not able to completely shut down accretion within the maximum allowed time $\tau_{\rm max} = 15$~Myr, because there is no density threshold that is assigned to turn off star formation. So, even if the feedback from a cluster particle has already blown out majority of the nearby molecular gas, there is still a trickle of mass that can be added to the cluster particle.

A more useful estimate of the formation timescale is the mass-weighted value, $\tau_{\rm ave}$, which contains information on the mass accretion history of an individual cluster. The right panel of Figure~\ref{fig:cluster_time} shows that the timescales of TURBSF and TURBSF2 models are much shorter, and more than 50\% of the cluster particles assemble their masses within 5 Myr. For the fiducial run, most of the cluster particles have $\tau_{\rm ave}$ smaller than 4 Myr, consistent with observations.

An important feature, illustrated by the plot, is a systematic dependence of the cluster formation timescale on the star formation efficiency and strength of stellar feedback. It is clear in both panels that, the higher $\epsilon_{\rm ff}$ the faster the clusters are formed. Therefore, comparison of the formation timescale of model clusters with the observed age spread of young star clusters would provide additional constraints on the local efficiency of star formation and feedback schemes.

\begin{figure}[t]
  \vspace{0cm}
\includegraphics[width=1.0\hsize]{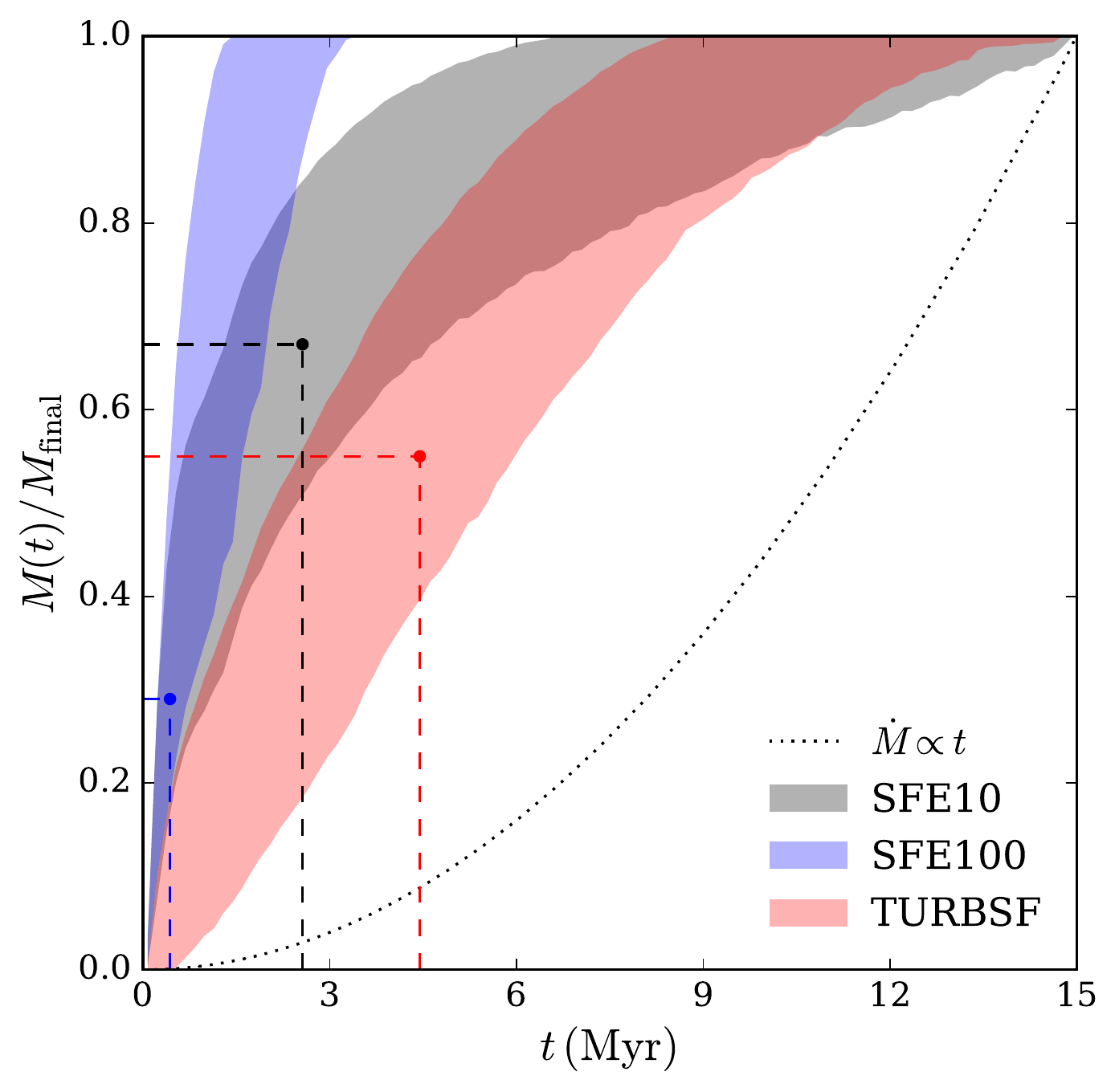}
  \vspace{-0.2cm}
\caption{\small The quartile (25-75 percentile) ranges of mass growth history for active clusters in SFE10 (black), SFE100 (blue), and TURBSF (red) runs, respectively. Median mass-averaged cluster formation timescales, $\tau_{\rm ave}$, for each model are shown as vertical dashed lines. If the mass growth of all clusters was exactly linear, $\dot{M} = \mathrm{const}$, these lines would intersect the median mass track at $M(\tau_{\rm ave})/M_{\rm final} = \onehalf$. The actual points at which they intersect vary from 0.29 to 0.67. A linearly increasing mass growth history, $\dot{M}\propto t$, predicted by \citet{murray_chang15} for the collapse of self-gravitating turbulent cloud is overplotted for comparison.}
  \vspace{0.2cm}
  \label{fig:cfh}
\end{figure}

Although it is impractical to store mass increments of all cluster particles at every local timestep, $\sim10^3$ yr, we randomly select 10\% of active clusters at the last epoch and output their mass growth history $M(t)$. The goal is to check whether the conclusions drawn from the distributions of $\tau_{\rm dur}$ and $\tau_{\rm ave}$ are consistent with this detailed output. We normalize the mass evolution of each cluster by its final mass $M_{\rm final}$, split $\tau_{\rm max}$ into 100 intervals of equal duration, and calculate the $25^{\rm th}$ and $75^{\rm th}$ percentiles of $M(t)/M_{\rm final}$ at each interval. These quartile ranges for models SFE10, SFE100, and TURBSF, are shown in Figure~\ref{fig:cfh}. We find that most of the clusters reach their maximum mass within 15 Myr. Especially, in SFE100 runs, clusters finish accretion in only 3 Myr. This is consistent with the distribution of $\tau_{\rm dur}$ in the left panel of Figure~\ref{fig:cluster_time}, and suggests that feedback from young stars extinguishes star formation effectively. We also find decreasing cluster formation timescale with increasing star formation efficiency, which is again consistent with the conclusion drawn from the distribution of $\tau_{\rm ave}$.

For all three models, the mass histories correspond to either constant or decreasing growth rate. None of them shows a linearly increasing rate, $\dot{M}\propto t$, as suggested by \citet{murray_chang15} for the collapse of a self-gravitating cloud. This is likely because stellar feedback slows down gas accretion as the mass of the cluster increases.

The overall shape of the mass history distribution appears to be linear for the most part, with a slower-increasing tail over the last few Myr. We can quantitative assess how close $\dot{M}$ is to a constant. If the growth rate of each cluster in a given run was truly linear, $\dot{M} = \mathrm{const}$, then it would reach its final mass in a time $t_{\rm final} = M_{\rm final}/\dot{M}$. The median time for the whole sample, $t_{\rm final}^{\rm med}$, would then be equivalent to the median duration timescale $\tau_{\rm dur}$ (because of linearity of $M(t)$), which in turn would be twice the mass-weighted timescale: $\tau_{\rm dur} = 2\tau_{\rm ave}$. The median mass accumulated within $\tau_{\rm ave}$ would be half of the final mass.  The values of $t_{\rm final}^{\rm med}$ or $\tau_{\rm ave}$ would differ from model to model, according to the trend with star formation efficiency and feedback. Figure~\ref{fig:cfh} demonstrates that the actual median mass $M(\tau_{\rm ave})$ is not exactly half of $M_{\rm final}$, but is not too different from it, for the three models shown.  Thus, clusters spend most of their formation period accreting at a steady rate.

From the large variation of the cluster formation history shown in Figure~\ref{fig:cfh}, we conclude that different clusters have dramatically different growth history during their active period. This reflects the intrinsic variation of the physical conditions in different star-forming regions, as well as the complex interplay between gas accretion and stellar feedback. This variation is also reflected in the large scatter of the resulting star formation efficiency, as we will discuss in Section~\ref{sec:var_eff}.

\section{Discussion}\label{sec:discussion}

\subsection{Comparison with other implementations of star formation}

Most cosmological simulations model star formation in a way that is similar to \citet{katz92}: a fraction of cold and dense gas is converted to star particles with a constant efficiency.
Some simulations adopt a fixed star particle mass and create particles with a Poisson random process at each timestep, while others use a relatively long sampling timescale (typically several Myr) and create star particles with varying masses.
One of the few alternatives, \citet{cen_ostriker_92} grow star particles over the galaxy dynamical time, according to a priori chosen star formation law. Moreover, in all of the above methods, star particle masses are determined before their feedback processes begin to influence the ambient environment. In contrast, in our model, cluster particles grow their mass continuously at each local timestep until their own feedback terminates the star formation episode. The mass accumulation history is resolved by a large number of timesteps, and the final particle mass is obtained self-consistently so that it can be considered a good proxy to the mass of a given cluster formation region.

In contrast to full galactic simulations, small-scale simulations of star and planet formation create collisionless sink particles that accrete matter over time. This approach has been used both in SPH \citep[e.g.][]{bate_etal95} and grid codes \citep{krumholz_etal04, federrath_etal10}. Sink particles are created in local density peaks if the gas is Jeans-unstable and gravitationally bound. The particles continue to accrete gas as long as these criteria are satisfied, which leads to a significant fraction of all gas in the simulation box to be converted into stars. Such simulations model small gas clouds, typically under $10^3\, M_\odot$, which do not produce many massive stars. Therefore, even when feedback is included, it does not drive strong outflows that would terminate a star formation episode. In this regard, our simulations differ as they model continuous inflow of gas onto large clouds, which generate many massive stars with stronger feedback.

Also, the sink particle approach usually requires that the accretion region be resolved by many cells, in order to determine the flow convergence and mass increment at each timestep. In a cosmological simulation, however, resolving cluster-forming regions requires a sub-parsec cell size, which is still computationally challenging. Instead, stellar particles in our model grow over a period of time, parametrized by $\epsilon_{\rm ff}$ and based on observations of individual GMCs on scales comparable to our spatial resolution \citep{krumholz_etal12}.

\subsection{On the shape of CIMF}\label{sec:CIMF_origin}

\begin{figure}[t]
  \vspace{0cm}
\includegraphics[width=1.0\hsize]{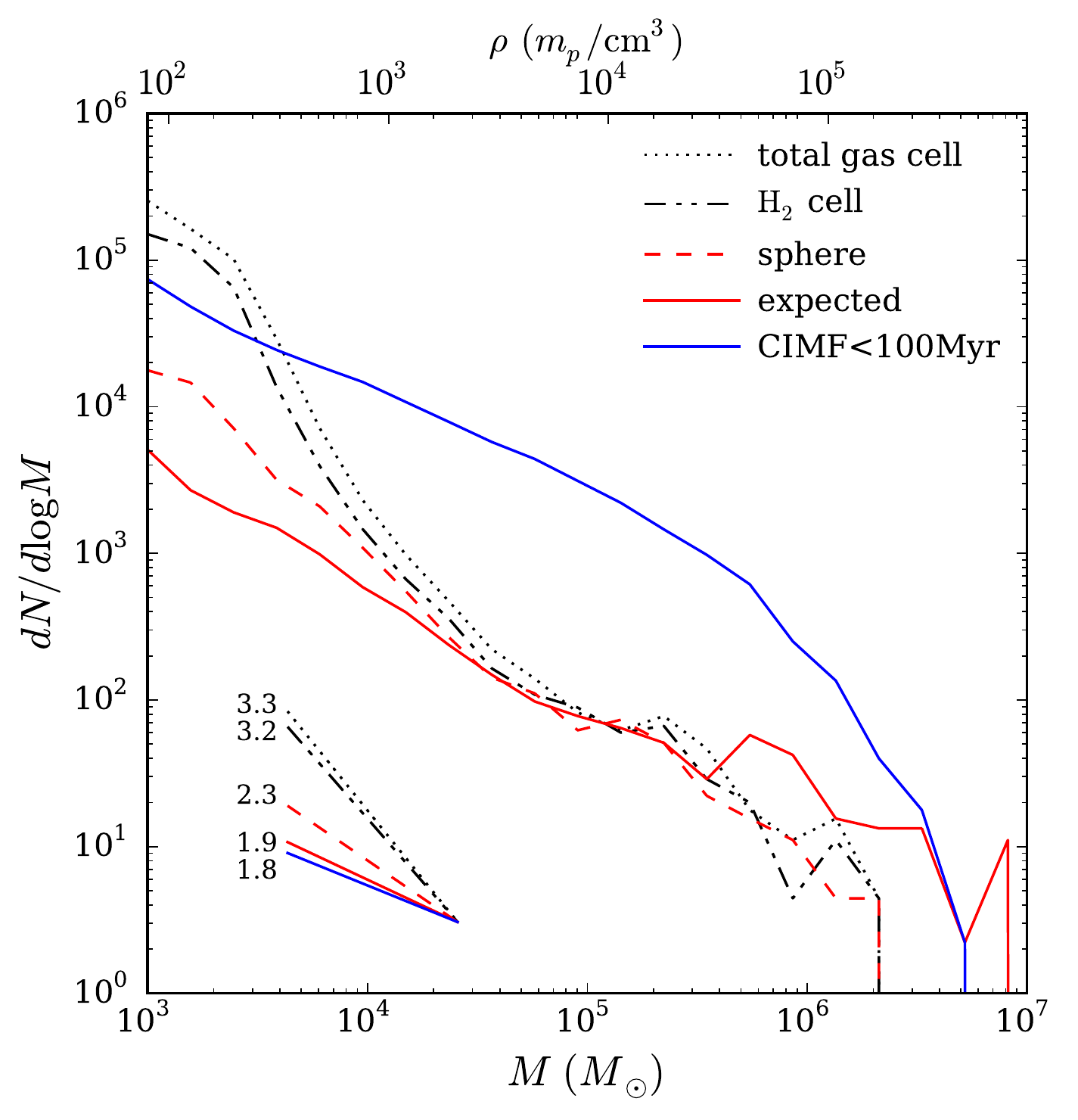}
  \vspace{-0.2cm}
\caption{\small Mass functions of gas cells (black), cluster-forming spheres (red), and star clusters (blue) within the virial radius of the main galaxy at $z\approx3.3$. Dotted and dashed-dotted black lines show the distribution of total and molecular gas mass in simulation cells, rescaled to the volume of the cluster-forming sphere.  Red dashed line is the mass function of molecular gas within the spheres. Red solid line shows the expected distribution of cluster mass from a simple growth model discussed in Section~\ref{sec:CIMF_origin}. Blue solid line shows CIMF for all clusters younger than 100 Myr. We choose this longer time interval to accumulate sufficient cluster number to characterize CIMF, but this means the normalization of CIMF differs from the other displayed mass functions. The plot illustrates only the differences in the shape of the mass functions. Dynamical disruption of clusters would also reduce the normalization of CIMF. The best-fit slopes of all mass functions in the range $10^3 - 10^5\, M_\odot$ are shown in the lower left corner with the corresponding line styles.}
  \vspace{0.2cm}
  \label{fig:MF_cell_sphere}
\end{figure}

In Section~\ref{sec:CIMF}, we examined the model CIMF in different galaxies at various epochs and found that it can be well described by the Schechter function.  The cutoff at high mass is related to the overall amount of cold dense gas available for star formation and therefore depends on the environment and evolutionary stage of the host galaxy.  The power-law slope at low mass is much more stable, with only a limited range of variation between $\alpha \approx 1.8-2.0$.  It is encouraging that this range is consistent with observations of young star clusters.  We can now investigate the physical origin of this CIMF slope within our model, and the connection between the cluster mass and physical properties of the ISM.

We calculate the molecular gas mass within each cluster-forming sphere by summing over the contributions of the central cell as well as the 26 neighbors: $M_{\rm sp} = \Sigma f_{{\rm sp},i}V_i f_{{\rm H_2},i}\rho_{{\rm gas},i}$. We also estimate the mass function of individual gas cells to compare it with the distribution of $M_{\rm sp}$. Since all clusters in our model accrete gas from spheres of fixed volume $V_{\rm sp}=\pi D_{\rm GMC}^3/6$, to make a fair comparison, we rescale the total (or molecular) cell mass to match the same volume: $\rho_{\rm gas}V_{\rm sp}$ (or $f_{\rm H_2}\rho_{\rm gas}V_{\rm sp}$).

Figure~\ref{fig:MF_cell_sphere} shows that the distribution of the rescaled total and molecular cell mass exhibits a steep slope at low masses, $\alpha > 3$. In contrast, the mass function of $M_{\rm sp}$ has a slope closer to that of the CIMF. This implies that our method in which the sphere encloses gas cells with various density is a more realistic way of modeling star cluster formation, than using simply the properties of the central cell.

However, the power-law slope of the mass function of $M_{\rm sp}$ is still steeper than $\alpha=2$ and the maximum sphere mass is only $\sim 2\times 10^6\, M_\odot$, several times smaller than the most massive model clusters. This discrepancy may stem from the fact that the instantaneous mass of the cluster-forming sphere does not contain the information of the evolving accretion rate due to feedback.

In order to connect the mass function of the spheres and CIMF, we present a simple cluster formation model by assuming a power-law mass accumulation history: $\dot{M}(t)=\dot{M}_0 \, (t/t_0)^m$, where $t_0$ is the time over which a cluster maintains the initial growth rate $\dot{M}_0=\epsilon_{\rm ff} M_{\rm sp}/\tau_{\rm ff}$.
The expected cluster mass in this model is:
\begin{equation}\label{eq:Msp_Mc}
M = \int_0^{\tau_{\rm max}} \dot{M}(t) \, dt = \frac{8\epsilon_{\rm ff} \, G^{1/2}}{\pi \, D_{\rm GMC}^{3/2}} \frac{\tau_{\rm max}^{m+1}}{(m+1) \, t_0^m} \, M_{\rm sp}^{3/2}.
\end{equation}
The value of the index $m$ does not affect the scaling with the initial sphere mass or density. Instead, the scaling $M \propto M_{\rm sp}^{3/2}$ reflects the dependence of the star formation rate on the free-fall time.

In Figure~\ref{fig:cfh}, we find that the median cluster formation history in the SFE10 run is generally described by a decreasing growth rate, roughly as $\dot{M}\propto t^{-0.5}$. We estimate the distribution of the expected cluster mass from the mass function of $M_{\rm sp}$ using Eq.~(\ref{eq:Msp_Mc}), with $m=-0.5$, $\epsilon_{\rm ff}=0.1$, $D_{\rm GMC}=10$ pc, $t_0=1$ Myr, and $\tau_{\rm max}=15$ Myr. Other choices of the index $m$ would only slightly affect the normalization of the mass function, but not its slope. We find that the slope of the expected cluster mass function ($\alpha \approx 1.9$) is very similar to that of the CIMF. Also the maximum expected mass can reach as high as $10^7 \, M_\odot$, similar to the maximum cluster mass in the simulations. This consistency indicates that the origin of CIMF is a combination of the mass function of GMCs and the accretion history within the cluster-forming regions.

In the expected cluster mass function, the power-law shape extends to very high mass and there exists no clear cutoff as seen in the CIMF. The high mass cutoff in the simulation may be caused by strong feedback from the most massive cluster particles. This feedback may result in more variation of the growth history than assumed in our simple estimate.

\subsection{On the similarities and differences with the constant-density-threshold model}

\citet{kravtsov_gnedin05} proposed a different scheme for forming massive star clusters, using post-processing of cosmological simulations. Those simulations were run with an earlier version of the ART code, with a traditional star formation recipe and a relative low threshold density for creating stellar particles. Even though the spatial resolution was 9 times lower than in our present simulations, it was sufficient to resolve the structure of largest gas clouds within a galactic disk at the same epoch, $z\approx 3.3$. The pressure within the clouds was dominated by turbulent motions, with an approximately flat velocity dispersion profile. This means that although the clouds were not strictly isothermal because the thermal pressure was negligible compared to the turbulence, they still had roughly $\rho(r) \propto r^{-2}$ density profiles. Adopting these profiles for the unresolved inner structure of the clouds, \citet{kravtsov_gnedin05} suggested that the central regions above a particular very high density could be forming massive star clusters. In order to produce a gravitationally bound cluster, the fraction of the dense gas turning into stars had to be above 50\% (and was taken to be 60\%); the rest of the gas is likely to be quickly blown out of the region by the winds and radiation of the young stars. The remaining stars would expand to a new hydrostatic equilibrium, and the final half-mass radius of the cluster $R_h$ could be estimated as the size of the region above the threshold density, corrected for the expansion. The value for the high density threshold ($\rho_{\rm CSF} = 10^4\, M_\odot\, \mathrm{pc}^{-3}$) was taken such that after the expansion, the final half-mass density would match the median observed density of Galactic globular clusters ($\approx 3\times 10^3\, M_\odot\, \mathrm{pc}^{-3}$). In that sense, the model had no free adjustable parameters.

Analysis of several outputs of that simulation showed a consistent shape of CIMF, similar to a power law with $\alpha \approx 2$, and also a sharper cutoff at high mass. An equally good fit was provided by a log-normal function.

What is similar and what is different in that model and our new model? In terms of technical execution, the old model was simpler, implemented on a few discrete outputs rather than in run-time, and based on a less sophisticated simulation. These differences aside, below we focus on the analysis of the results of the two models, as they relate to the origin of CIMF.

In the old model, the constant density threshold implies that the average density, $M/R_h^3$, is the same for all model clusters. The half-mass radius is set by the condition $\rho(R_h) = \rho_{\rm CSF}= \mathrm{const}$, or $\rho_{\rm cell} (L_{\rm cell}/R_h)^2 = \mathrm{const}$, which means it scales as $R_h \propto \rho_{\rm cell}^{1/2}$, and the cluster mass scales as $M \propto \rho_{\rm cell}^{3/2}$. In the new model, the cluster mass scales with cell density in exactly the same way, $M \propto \rho^{3/2}$, but for a different reason -- because of the dependence of the SFR on the free-fall time. Of course, the time evolution of the cluster accretion rate in the continuous model can modify this scaling, but the overall dependence on cell density should still be similar in the two models. Thus, if the distribution of cloud density was the same in both simulations, the two models would result in the same CIMF. 

At the highest densities in the old simulation, the probability distribution function could be fit by a single power law, ${dN/d\log{\rho}} \propto \rho^{-n}$, with $n \approx 1.4$. Given the scaling of cluster mass on density described above, this results in the CIMF obeying a power law ${dN/d\log{M}} \propto M^{-1-2n/3}$. The value of the slope $\alpha = 1+2n/3 \approx 2$.

In the new simulations the slope of the density PDF is steeper and changes, as shown by Figure~\ref{fig:MF_cell_sphere}, from $n \approx 3.3$ for $\rho \sim 10^3\, m_p$~cm$^{-3}$ to $n \approx 2$ for $\rho > 10^4\, m_p$~cm$^{-3}$.  The ART code now incorporates substantially revised physics of gas heating and cooling and includes transfer of ionizing radiation through the ISM. Thus the differences in the density PDF are not surprising. Applying the old model to the new simulations would result in a steeper CIMF ($\alpha \approx 2.3-3.2$). This again emphasizes the need for a continuous model of cluster accretion that we developed in this paper.

The advantage of the \citet{kravtsov_gnedin05} model was the ability to estimate cluster size, in addition to mass. In our new simulations we can only set an upper limit on the cluster size being smaller than the radius of the sphere, $D_{\rm GMC}/2$. With this limit, we can estimate the lower limit on the average cluster density in our continuous model,
$$
   \rho_{\rm av} = {M \over 4\pi/3 \, (D_{\rm GMC}/2)^3} \propto \rho^{3/2},
$$
and compare it with the central cell density. Calculating all the coefficients results in
$$
   {\rho_{\rm av} \over \rho} \approx 0.48 \left({\epsilon_{\rm ff} \over 0.1}\right)\left({\tau_{\rm max} \, t_0 \over 15\, \mathrm{Myr}^2}\right) \left({\rho \over 10^3 \, m_p \,\mathrm{cm}^{-3}}\right)^{1/2}.
$$
At low densities, this estimated $\rho_{\rm av}$ can be lower than the cell density, because the local efficiency of star formation $\epsilon_{\rm ff}$ is low and only fraction of the gas is being converted into stars. Conversely, at densities above $\rho \sim 4\times 10^3\, m_p \,\mathrm{cm}^{-3}$, the cluster accretes enough material to compensate for low $\epsilon_{\rm ff}$ and becomes dense than the initial cloud. Note that the actual half-mass density may be a factor of few higher than our estimate, because of additional dissipation on sub-grid scales during cluster formation.

\subsection{On the high mass end of CIMF}

Although various observations of young massive clusters in nearby galaxies suggest a roughly power-law mass function, it is still debated whether there exists a high mass cutoff in the CIMF and whether the cutoff mass is environment-dependent.\citet{larsen_02} found a correlation between SFR and the V-band magnitude of the brightest clusters and suggested that cluster formation is a stochastic process in which the maximum cluster luminosity can be explained by statistical sampling from a universal CIMF, the so-called ``size-of-the-sample'' effect. Other observations in various environments suggest that the variation of the high mass end of the CIMF may reflect different physical condition of their host galaxies \citep[e.g.][]{bastian_08,goddard_etal10, adamo_etal15}. More recently, \citet{sun_etal16} analyzed the cluster mass-galactocentric distance relation in four galaxies and concluded that, at least statistically, it is hard to rule out the environment-independent cluster formation scenario.

In our simulations, we have a large number of cluster samples formed at different epochs in different galaxies. Therefore, we can study the influence of environment on the model CIMF. First, we find a strong correlation between the fraction of massive clusters $\Gamma_{\rm sim}$ and the SFR surface density, $\Sigma_{\rm SFR}$, shown in Figure~\ref{fig:sfr_cfe}. If the shape of the CIMF does not change with environment and the most massive clusters are formed merely due to the ``size-of-the-sample'' effect,  average $\Gamma_{\rm sim}$ will be constant for different SFR and have larger scatter for lower SFR cases. However, the apparent positive correlation between $\Gamma_{\rm sim}$ and $\Sigma_{\rm SFR}$ suggests that massive clusters are preferentially formed in starbursts. This in turn implies an environment-dependent cluster formation process and disfavors the stochastic scenario. Second, there exists a positive correlation between the cutoff mass $M_{\rm cut}$ and SFR of the host galaxy, which directly reveals environmental dependency of CIMF. The maximum cluster mass would be largely overestimated by assuming a pure power-law CIMF, but it is in agreement with the predicted value when assuming a Schechter function with $M_{\rm cut}$ that is inferred from the empirical $M_{\rm cut}$-SFR relation. All these results validate a Schechter function fit with the environment-dependent cutoff mass as best description of CIMF.

\subsection{On the variation of star formation efficiency} \label{sec:var_eff}

\begin{figure}[t]
  \vspace{0cm}
\includegraphics[width=1.0\hsize]{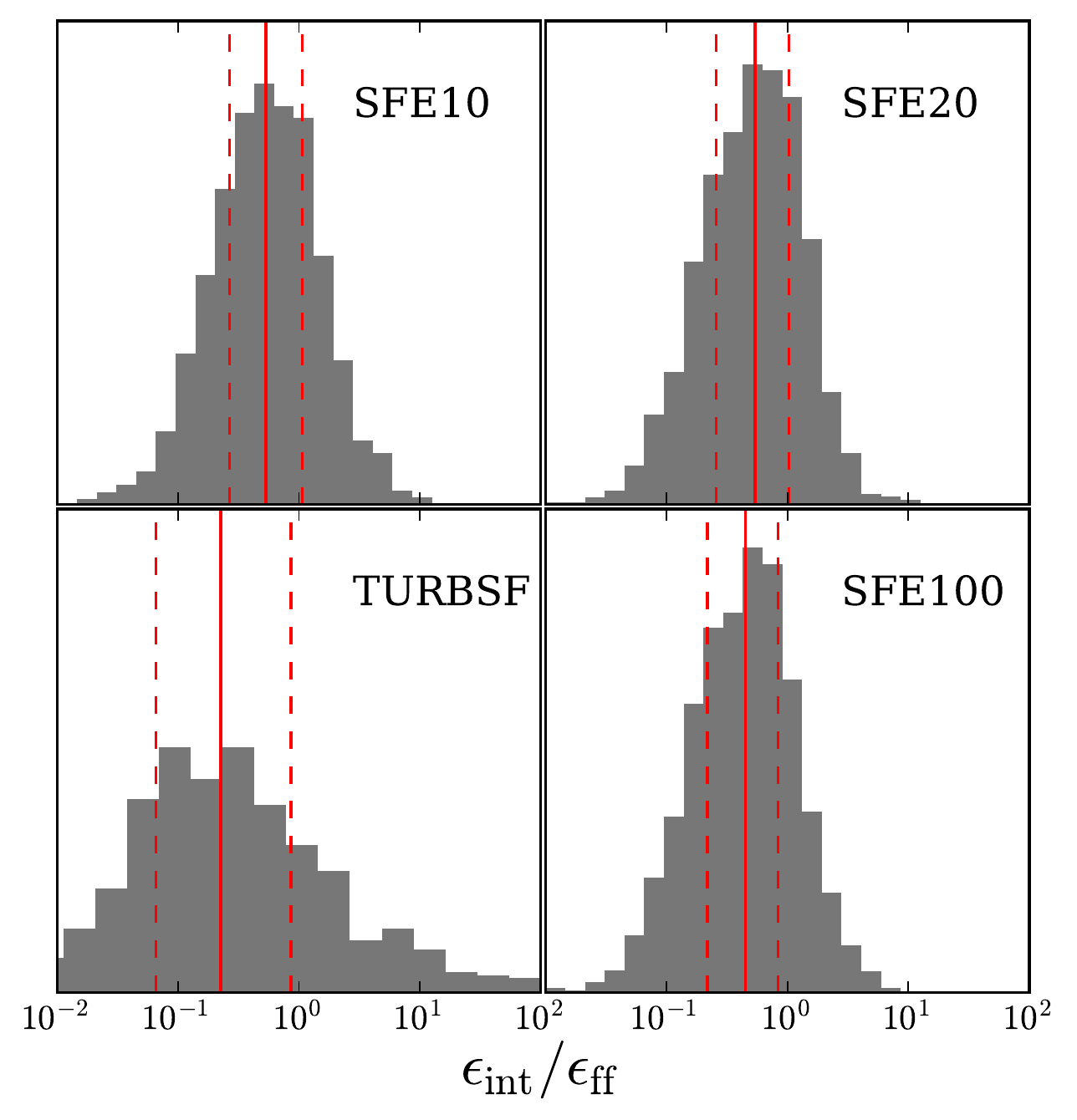}
  \vspace{-0.2cm}
\caption{\small Distribution of the integral star formation efficiency $\epsilon_{\rm int}$ for four models with different local efficiency $\epsilon_{\rm ff}$. We show the ratio $\epsilon_{\rm int}/\epsilon_{\rm ff}$ to emphasize the spread of values resulting from different accretion histories. Vertical lines represent the median and the 25-75 percentile range for each distribution.}
  \vspace{0.3cm}
  \label{fig:sfe}
\end{figure}

Recent observations of star-forming complexes suggest that the star formation efficiency on GMC scales varies significantly over 2-3 orders of magnitude \citep[e.g.][]{lada_etal10, murray11, evans_etal14,vutisalchavakul_etal16}. Such a large scatter may come from either the variation in cloud properties \citep[e.g.][]{krumholz_mckee05, padoan_nordlund11, federrath_klessen12, hennebelle_chabrier13}, or the time-evolution of self-gravitating clouds \citep{feldmann_gnedin11, lee_etal15}, or both. Realistic modeling of star formation in galaxy formation simulations should be able to reproduce this range of variation. Here we examine how the differences in the formation history of each model cluster affect the value of the efficiency that would be inferred for it  at the time it begins star formation.

We define the integral efficiency $\epsilon_{\rm int}$ such that the final cluster mass, $M$, can be obtained by a constant growth rate $\epsilon_{\rm int} M_{\rm sp, 0}/ \tau_{\rm ff,0}$ over the mass-weighted cluster formation timescale $\tau_{\rm ave}$:
\begin{equation}
  \epsilon_{\rm int} \equiv \frac{ M /\tau_{\rm ave} } { M_{\rm sp, 0} /\tau_{\rm ff,0}}.
  \label{eq:epsint}
\end{equation}
Here $M_{\rm sp,0}$ is the initial mass of the cluster-forming sphere, and $\tau_{\rm ff,0}$ is the free-fall time when the cluster particle is created.

Figure~\ref{fig:sfe} shows the distribution of $\epsilon_{\rm int}$, normalized by the intrinsic star formation efficiency $\epsilon_{\rm ff}$, for both fixed $\epsilon_{\rm ff}$ and turbulence-based $\epsilon_{\rm ff}$ runs. A mass-weighted average $\epsilon_{\rm ff}\approx 1.4\%$ is used for TURBSF run. The median values of $\epsilon_{\rm int}/\epsilon_{\rm ff}$ for all models are smaller than unity. This is possibly due to the outflows from the cluster-forming region caused by stellar feedback, which reduce the amount of gas available for star formation. More importantly, we find a large variation of $\epsilon_{\rm int}$ around the median. For the fixed $\epsilon_{\rm ff}$ runs (SFE10, SFE20, SFE100), $\epsilon_{\rm int}$ varies over more than 2 orders of magnitude. This variation is a natural consequence of the evolution of properties of cluster-forming spheres caused by the continuous gas accretion, cluster formation, and stellar feedback. The variation is even larger in TURBSF run, where $\epsilon_{\rm ff}$ is re-evaluated at each timestep during the cluster formation period. Although we cannot compare distributions of efficiencies in our simulations and observations directly, the large variation shown by the models in Figure~\ref{fig:sfe} is in good qualitative agreement with observational estimates. This demonstrates that modeling continuous formation of star cluster particles in galaxy formation simulations results in a more realistic distribution of cluster masses.

\subsection{On the origin of globular clusters}

Very massive star clusters, with $M\gtrsim 2\times 10^5\, M_\odot$, have lifetime comparable to the age of the Universe \citep[e.g.,][]{muratov_gnedin10}. Therefore, they can be considered as progenitors of present-day globular clusters (GCs). Studying the formation of these clusters at high redshift gives us an opportunity to investigate the origin of GC populations. For example, \citet{ashman_zepf92} proposed a model of GC formation in the gas-rich major mergers. \citet{muratov_gnedin10} and \citet{li_gnedin14} incorporated this scenario in the semi-analytical model to study GC properties in the Milky Way and the Virgo Cluster galaxies, and successfully reproduced the multi-modal metallicity distributions of their GC systems.

Following the idea that cluster formation is a strong environment-dependent process, there are two requirements for the formation of massive clusters. First, the CIMF needs to have a high cutoff mass so that the probability of sampling massive clusters is not too low. For instance, in order to obtain $M_{\rm cut} = 10^6\, M_\odot$, the corresponding SFR needs to reach as high as $14\, M_\odot\, {\rm yr}^{-1}$, based on Eq.~(\ref{eq:Mcut_SFR}). The SFR of this level is hard to achieve during the quiescent evolutionary stage of a Milky Way-sized galaxy, but it can be reached when the galaxy experiences major mergers. Second, to form such massive clusters, the host galaxy needs to have a large reservoir of cold gas. Indeed, as we can see from Section \ref{sec:merger}, massive clusters in our simulations are preferentially formed during the gas-rich major merger events.

Another possible dynamical effect of major mergers on GC formation is that the merger kicks massive clusters out of their birthplace in the galactic disk with high gas and stellar density, and places them into the halo where the tidal field is weak. This process helps to protect the newly formed clusters from being quickly destroyed by the intense tidal field of the disk. Such a two-stage formation and evolution scenario \citep{kravtsov_gnedin05, kruijssen15} can be tested in our model by analyzing the evolution of cluster bound fraction, as discussed in Section~\ref{sec:disruption}. In a follow-up paper we will investigate the transition from young massive clusters at high redshift to the GCs that we observe in the local Universe.

\section{Summary}\label{sec:sum}

We introduced a new star formation implementation, in which star cluster is considered a unit of star formation and cluster particles accumulate mass from a fixed sphere similar in size to GMC clumps. The mass growth of a given particle is terminated by its own feedback, so that its final mass is obtained self-consistently and represents the actual mass of a newly formed star cluster within the GMC. We implemented this cluster formation model in the ART code, and performed several high-resolution cosmological simulations of a Milky Way-sized galaxy under different model parameters. We analyzed the properties of young massive clusters in the simulated galaxies at high redshift ($z>3$) and the main results are summarized below.

\begin{itemize}
\item The CIMF is best described by a Schechter function, with a power-law slope $\alpha\approx1.8$, in agreement with observations of young massive clusters in nearby galaxies. The shape of the CIMF is not sensitive to the model parameters, except for model SFE100. The CIMF of SFE100, with 100\% star formation efficiency per free-fall time, has a cutoff mass that is much smaller than the other models and is inconsistent with observations.

\item We find a positive correlation between the SFR and the cutoff mass: $M_{\rm cut} \propto \mathrm{SFR}^{1.6}$. The maximum cluster mass also increases with the SFR, $M_{\rm max} \propto \mathrm{SFR}^{1.4}$, consistent with the value expected for a Schechter-like CIMF. We also find a tight correlation between the SFR surface density and the fraction of star formation contained in massive star clusters. The maximum mass decreases with distance from the galaxy center, as the SFR density decreases. All these trends suggest that cluster formation is a local process that strongly depends on the galactic environment. The scenario that clusters are formed solely by stochastic sampling from a universal CIMF is ruled out.

\item Maximum cluster mass in a given galaxy reaches $10^7\, M_\odot$ when SFR$\,\ga 10\, M_\odot\, {\rm yr}^{-1}$, but falls below $10^6\, M_\odot$ when SFR$\,\la 3\ M_\odot\,{\rm yr}^{-1}$.

\item Feedback from young stars extinguishes star formation in a GMC within 4~Myr, consistent with the observed age spread of young star clusters. The cluster formation timescale decreases systematically with increasing local efficiency of star formation, $\epsilon_{\rm ff}$, to a minimum of 0.5~Myr. This systematic trend indicates that cluster formation is strongly influenced by its own feedback. The power-law slope of the CIMF arises from a combination of the mass function of GMC clumps and the feedback-regulated mass accretion history.

\item We show that our cluster formation model leads to a large variation of the integral star formation efficiency (Eq.~\ref{eq:epsint}) even when $\epsilon_{\rm ff}$ is kept constant. The range of this variation spans about two orders of magnitude, similar to recent observations of star formation efficiency within GMCs. It is a natural consequence of the evolution of gas density in star-forming regions, caused by continuous gas infall, cluster formation, and stellar feedback.
 \newline

\end{itemize}

\vspace{5mm}

\acknowledgements
The simulations used in this work have been performed on the Joint Fermilab-KICP cluster ``Fulla'' at Fermilab. We thank Greg Bryan, Brian O'Shea, Andreas Burkert, Cliff Johnson, Alexander Muratov, and Norman Murray for helpful discussions. We are grateful to Angela Adamo for sending us their data, and Peter Behroozi for helping us construct merger trees using the ``consistent tree'' algorithm. H.L. and O.G. were supported in part by NASA through grant NNX12AG44G, and by NSF through grant AST-1412144. This work was begun during the "Gravity's Loyal Opposition: The Physics of Star Formation Feedback" workshop at the Kavli Institute for Theoretical Physics in Santa Barbara, which is supported in part by the National Science Foundation under grant PHY11-25915.

\appendix

Young stars in massive clusters produce copious amount of ionizing radiation that travels through the ISM and may potentially leave the host galaxy.  The escaped radiation would contribute to the extragalactic UV flux that reionized the universe by $z\approx6$ \citep[e.g.][]{gnedin16}. Before this radiation escapes the galaxy, a significant portion will be absorbed by neutral hydrogen located close to the stellar sources.  Since the actual sizes of stars cannot be resolved in a cosmological simulation, calculation of the escape fraction depends on the numerical resolution and the modeled structure of the neutral ISM.  The better the resolution, the more accurately the escaped fraction can be calculated.  Our simulations reach smaller cell size than most current galactic simulations, and therefore, can provide a useful estimate of the escape fraction in the inner few hundred parsecs.

We calculate the absorption by neutral hydrogen using the absorption cross-section integrated over the radiation spectrum, separately for atomic hydrogen and molecular hydrogen.  These cross-sections are about 3 times lower than the monochromatic cross-sections at the ionization edge: 
$$
  \langle\sigma_{\rm HI}\rangle \equiv {\int \sigma_{\rm HI}(\nu) \, f_\nu \, d\nu \over \int f_\nu \, d\nu} = 0.34 \times \sigma_{\rm HI}(13.6\, \mathrm{eV})
$$
$$
  \langle\sigma_{\rm H_2}\rangle = 0.36 \times \sigma_{\rm H_2}(15.4\, \mathrm{eV}) 
$$
for the power-law spectrum $f_\nu \propto \nu^{-3}$.  We take $\sigma_{\rm HI}(13.6\, \mathrm{eV}) = 6.3 \times 10^{-18}\, {\rm cm}^2$ and $\sigma_{\rm H_2}(15.4\, \mathrm{eV}) = 10^{-17}\, {\rm cm}^2$. Most of molecular hydrogen around young star clusters may be first dissociated by the Lyman-Werner radiation, but any remaining molecules would absorb ionizing photons just like the atomic component. The total optical depth for absorption is $\langle\sigma_{\rm HI}\rangle \, N_{\rm HI} + \langle\sigma_{\rm H_2}\rangle \, N_{\rm H_2}$. 

We find very small escape fractions at the virial radius of our simulated galaxies, below 0.1\%, in the redshift range $z \approx 5-7$. Such low values may indicate that our implementation of stellar feedback is not sufficient to keep enough hydrogen ionized within the galactic disks. Alternatively, this result may indicate that relatively small galaxies, that we have in our simulation box, are inefficient at transporting their ionizing photons to the intergalactic medium. The range of stellar mass we probe at these epochs is $10^{8} - 10^{9}\, M_\odot$, with the maximum star formation rate of $2\, M_\odot/\mathrm{yr}$. More luminous galaxies may need to provide the majority of the ionizing flux.
We also find a large variation of the value of $f_{\rm esc}$ among simulated galaxies of the same mass, similar to the results of SPH simulations with similar spatial resolution by \citet{paardekooper_etal15}.  If this variation is confirmed by future studies, it would add a stochastic component to simple models of reionization.

Our simulations include the extragalactic background contribution to the UV flux \citep{haardt_madau01} which dominates the internal flux of young stars outside about 2 kpc from the galaxy center.  Therefore, the simulations are better suited for calculation of the escape fraction in the inner parts of the galaxies.  For example, we can estimate the amount of unresolved absorption in the large-scale CROC simulations of reionization \citep{gnedin14}, which use the smallest cell size of $875/(1+z)$ pc.  We calculated the azimuthally-averaged escape fractions of the sources in our fiducial model simulation, from our smallest cell to a distance corresponding to one CROC cell.  The escape fractions never exceed 1\%, which indicates that the absorption in the immediate environment of stellar sources is very strong.  However, these results should be interpreted with caution, because the radiative transfer routine in the ART code becomes accurate only on the scale of several cell sizes, and may over smooth the neutral gas density in close vicinity of the sources.  Also, if the stellar feedback is underestimated in our simulations, it would also make it harder for the ionizing radiation to escape along the holes in the ISM evacuated by gas outflows.  A task of accurate calculation of the transfer of ionizing flux within the galaxies is difficult and will require specialized small-scale simulations, beyond the scope of our current models.

\makeatletter\@chicagotrue\makeatother

\bibliographystyle{apj}
\bibliography{bulge}

\begin{thebibliography}{}
\expandafter\ifx\csname natexlab\endcsname\relax\def\natexlab#1{#1}\fi

\bibitem[{{Adamo} {et~al.}(2015){Adamo}, {Kruijssen}, {Bastian}, {Silva-Villa},
  \& {Ryon}}]{adamo_etal15}
{Adamo}, A., {Kruijssen}, J.~M.~D., {Bastian}, N., {Silva-Villa}, E., \&
  {Ryon}, J. 2015, \mnras, 452, 246

\bibitem[{{Agertz} \& {Kravtsov}(2015)}]{agertz_kravtsov_15}
{Agertz}, O., \& {Kravtsov}, A.~V. 2015, \apj, 804, 18

\bibitem[{{Agertz} \& {Kravtsov}(2016)}]{agertz_kravtsov_16}
---. 2016, \apj, 824, 79

\bibitem[{{Agertz} {et~al.}(2013){Agertz}, {Kravtsov}, {Leitner}, \&
  {Gnedin}}]{agertz_etal13}
{Agertz}, O., {Kravtsov}, A.~V., {Leitner}, S.~N., \& {Gnedin}, N.~Y. 2013,
  \apj, 770, 25

\bibitem[{{Ashman} \& {Zepf}(1992)}]{ashman_zepf92}
{Ashman}, K.~M., \& {Zepf}, S.~E. 1992, \apj, 384, 50

\bibitem[{{Bastian}(2008)}]{bastian_08}
{Bastian}, N. 2008, \mnras, 390, 759

\bibitem[{{Bate} {et~al.}(1995){Bate}, {Bonnell}, \& {Price}}]{bate_etal95}
{Bate}, M.~R., {Bonnell}, I.~A., \& {Price}, N.~M. 1995, \mnras, 277, 362

\bibitem[{{Behroozi} {et~al.}(2013{\natexlab{a}}){Behroozi}, {Wechsler}, \&
  {Conroy}}]{behroozi_etal13}
{Behroozi}, P.~S., {Wechsler}, R.~H., \& {Conroy}, C. 2013{\natexlab{a}}, \apj,
  770, 57

\bibitem[{{Behroozi} {et~al.}(2013{\natexlab{b}}){Behroozi}, {Wechsler}, \&
  {Wu}}]{behroozi_rockstar}
{Behroozi}, P.~S., {Wechsler}, R.~H., \& {Wu}, H.-Y. 2013{\natexlab{b}}, \apj,
  762, 109

\bibitem[{{Behroozi} {et~al.}(2013{\natexlab{c}}){Behroozi}, {Wechsler}, {Wu},
  {Busha}, {Klypin}, \& {Primack}}]{behroozi_tree}
{Behroozi}, P.~S., {Wechsler}, R.~H., {Wu}, H.-Y., {et~al.} 2013{\natexlab{c}},
  \apj, 763, 18

\bibitem[{{Behroozi} {et~al.}(2015){Behroozi}, {Zhu}, {Ferguson}, {Hearin},
  {Lotz}, {Silk}, {Kassin}, {Lu}, {Croton}, {Somerville}, \&
  {Watson}}]{behroozi_etal15}
{Behroozi}, P.~S., {Zhu}, G., {Ferguson}, H.~C., {et~al.} 2015, \mnras, 450,
  1546

\bibitem[{{Booth} {et~al.}(2013){Booth}, {Agertz}, {Kravtsov}, \&
  {Gnedin}}]{booth_etal13}
{Booth}, C.~M., {Agertz}, O., {Kravtsov}, A.~V., \& {Gnedin}, N.~Y. 2013,
  \apjl, 777, L16

\bibitem[{{Cen} \& {Ostriker}(1992)}]{cen_ostriker_92}
{Cen}, R., \& {Ostriker}, J.~P. 1992, \apjl, 399, L113

\bibitem[{{Ceverino} {et~al.}(2014){Ceverino}, {Klypin}, {Klimek},
  {Trujillo-Gomez}, {Churchill}, {Primack}, \& {Dekel}}]{ceverino_etal14}
{Ceverino}, D., {Klypin}, A., {Klimek}, E.~S., {et~al.} 2014, \mnras, 442, 1545

\bibitem[{{Evans} {et~al.}(2014){Evans}, {Heiderman}, \&
  {Vutisalchavakul}}]{evans_etal14}
{Evans}, II, N.~J., {Heiderman}, A., \& {Vutisalchavakul}, N. 2014, \apj, 782,
  114

\bibitem[{{Federrath}(2015)}]{federrath15}
{Federrath}, C. 2015, \mnras, 450, 4035

\bibitem[{{Federrath} {et~al.}(2010){Federrath}, {Banerjee}, {Clark}, \&
  {Klessen}}]{federrath_etal10}
{Federrath}, C., {Banerjee}, R., {Clark}, P.~C., \& {Klessen}, R.~S. 2010,
  \apj, 713, 269

\bibitem[{{Federrath} \& {Klessen}(2012)}]{federrath_klessen12}
{Federrath}, C., \& {Klessen}, R.~S. 2012, \apj, 761, 156

\bibitem[{{Feldmann} \& {Gnedin}(2011)}]{feldmann_gnedin11}
{Feldmann}, R., \& {Gnedin}, N.~Y. 2011, \apjl, 727, L12

\bibitem[{{Gieles} {et~al.}(2006){Gieles}, {Larsen}, {Bastian}, \&
  {Stein}}]{gieles_etal06}
{Gieles}, M., {Larsen}, S.~S., {Bastian}, N., \& {Stein}, I.~T. 2006, \aap,
  450, 129

\bibitem[{{Gnedin}(2014)}]{gnedin14}
{Gnedin}, N.~Y. 2014, \apj, 793, 29

\bibitem[{{Gnedin}(2016)}]{gnedin16}
---. 2016, ArXiv e-prints, arXiv:1603.07729

\bibitem[{{Gnedin} \& {Abel}(2001)}]{gnedin_abel01}
{Gnedin}, N.~Y., \& {Abel}, T. 2001, \na, 6, 437

\bibitem[{{Gnedin} \& {Kravtsov}(2011)}]{gnedin_kravtsov11}
{Gnedin}, N.~Y., \& {Kravtsov}, A.~V. 2011, \apj, 728, 88

\bibitem[{{Gnedin} {et~al.}(2011){Gnedin}, {Kravtsov}, \&
  {Rudd}}]{gnedin_etal11}
{Gnedin}, N.~Y., {Kravtsov}, A.~V., \& {Rudd}, D.~H. 2011, \apjs, 194, 46

\bibitem[{{Gnedin} {et~al.}(2014){Gnedin}, {Ostriker}, \&
  {Tremaine}}]{gnedin_etal14}
{Gnedin}, O.~Y., {Ostriker}, J.~P., \& {Tremaine}, S. 2014, \apj, 785, 71

\bibitem[{{Goddard} {et~al.}(2010){Goddard}, {Bastian}, \&
  {Kennicutt}}]{goddard_etal10}
{Goddard}, Q.~E., {Bastian}, N., \& {Kennicutt}, R.~C. 2010, \mnras, 405, 857

\bibitem[{{Governato} {et~al.}(2007){Governato}, {Willman}, {Mayer}, {Brooks},
  {Stinson}, {Valenzuela}, {Wadsley}, \& {Quinn}}]{governato_etal07}
{Governato}, F., {Willman}, B., {Mayer}, L., {et~al.} 2007, \mnras, 374, 1479

\bibitem[{{Governato} {et~al.}(2012){Governato}, {Zolotov}, {Pontzen},
  {Christensen}, {Oh}, {Brooks}, {Quinn}, {Shen}, \&
  {Wadsley}}]{governato_etal12}
{Governato}, F., {Zolotov}, A., {Pontzen}, A., {et~al.} 2012, \mnras, 422, 1231

\bibitem[{{Haardt} \& {Madau}(2001)}]{haardt_madau01}
{Haardt}, F., \& {Madau}, P. 2001, in Clusters of Galaxies and the High
  Redshift Universe Observed in X-rays, ed. D.~M. {Neumann} \& J.~T.~V. {Tran}

\bibitem[{{Hartmann} {et~al.}(2012){Hartmann}, {Ballesteros-Paredes}, \&
  {Heitsch}}]{hartmann_etal12}
{Hartmann}, L., {Ballesteros-Paredes}, J., \& {Heitsch}, F. 2012, \mnras, 420,
  1457

\bibitem[{{Hennebelle} \& {Chabrier}(2013)}]{hennebelle_chabrier13}
{Hennebelle}, P., \& {Chabrier}, G. 2013, \apj, 770, 150

\bibitem[{{Hollyhead} {et~al.}(2015){Hollyhead}, {Bastian}, {Adamo},
  {Silva-Villa}, {Dale}, {Ryon}, \& {Gazak}}]{hollyhead_etal15}
{Hollyhead}, K., {Bastian}, N., {Adamo}, A., {et~al.} 2015, \mnras, 449, 1106

\bibitem[{{Hopkins} {et~al.}(2014){Hopkins}, {Kere{\v s}}, {O{\~n}orbe},
  {Faucher-Gigu{\`e}re}, {Quataert}, {Murray}, \& {Bullock}}]{hopkins_etal14}
{Hopkins}, P.~F., {Kere{\v s}}, D., {O{\~n}orbe}, J., {et~al.} 2014, \mnras,
  445, 581

\bibitem[{{Hopkins} {et~al.}(2011){Hopkins}, {Quataert}, \&
  {Murray}}]{hopkins_etal11}
{Hopkins}, P.~F., {Quataert}, E., \& {Murray}, N. 2011, \mnras, 417, 950

\bibitem[{{Hummels} \& {Bryan}(2012)}]{hummels_bryan12}
{Hummels}, C.~B., \& {Bryan}, G.~L. 2012, \apj, 749, 140

\bibitem[{{Katz}(1992)}]{katz92}
{Katz}, N. 1992, \apj, 391, 502

\bibitem[{{Keller} {et~al.}(2015){Keller}, {Wadsley}, \&
  {Couchman}}]{keller_etal15}
{Keller}, B.~W., {Wadsley}, J., \& {Couchman}, H.~M.~P. 2015, \mnras, 453, 3499

\bibitem[{{Kravtsov}(1999)}]{kravtsov99}
{Kravtsov}, A.~V. 1999, PhD thesis, NEW MEXICO STATE UNIVERSITY

\bibitem[{{Kravtsov}(2003)}]{kravtsov03}
---. 2003, \apjl, 590, L1

\bibitem[{{Kravtsov} \& {Gnedin}(2005)}]{kravtsov_gnedin05}
{Kravtsov}, A.~V., \& {Gnedin}, O.~Y. 2005, \apj, 623, 650

\bibitem[{{Kravtsov} {et~al.}(1997){Kravtsov}, {Klypin}, \&
  {Khokhlov}}]{kravtsov_etal97}
{Kravtsov}, A.~V., {Klypin}, A.~A., \& {Khokhlov}, A.~M. 1997, \apjs, 111, 73

\bibitem[{{Kritsuk} {et~al.}(2013){Kritsuk}, {Lee}, \&
  {Norman}}]{kritsuk_etal13}
{Kritsuk}, A.~G., {Lee}, C.~T., \& {Norman}, M.~L. 2013, \mnras, 436, 3247

\bibitem[{{Kroupa}(2001)}]{kroupa_01}
{Kroupa}, P. 2001, \mnras, 322, 231

\bibitem[{{Kruijssen}(2014)}]{kruijssen14}
{Kruijssen}, J.~M.~D. 2014, Classical and Quantum Gravity, 31, 244006

\bibitem[{{Kruijssen}(2015)}]{kruijssen15}
---. 2015, \mnras, 454, 1658

\bibitem[{{Krumholz} {et~al.}(2012){Krumholz}, {Dekel}, \&
  {McKee}}]{krumholz_etal12}
{Krumholz}, M.~R., {Dekel}, A., \& {McKee}, C.~F. 2012, \apj, 745, 69

\bibitem[{{Krumholz} \& {McKee}(2005)}]{krumholz_mckee05}
{Krumholz}, M.~R., \& {McKee}, C.~F. 2005, \apj, 630, 250

\bibitem[{{Krumholz} {et~al.}(2004){Krumholz}, {McKee}, \&
  {Klein}}]{krumholz_etal04}
{Krumholz}, M.~R., {McKee}, C.~F., \& {Klein}, R.~I. 2004, in Astronomical
  Society of the Pacific Conference Series, Vol. 323, Star Formation in the
  Interstellar Medium: In Honor of David Hollenbach, ed. D.~{Johnstone}, F.~C.
  {Adams}, D.~N.~C. {Lin}, D.~A. {Neufeeld}, \& E.~C. {Ostriker}, 401

\bibitem[{{Lada} \& {Lada}(2003)}]{lada_lada03}
{Lada}, C.~J., \& {Lada}, E.~A. 2003, \araa, 41, 57

\bibitem[{{Lada} {et~al.}(2010){Lada}, {Lombardi}, \& {Alves}}]{lada_etal10}
{Lada}, C.~J., {Lombardi}, M., \& {Alves}, J.~F. 2010, \apj, 724, 687

\bibitem[{{Larsen}(2002)}]{larsen_02}
{Larsen}, S.~S. 2002, \aj, 124, 1393

\bibitem[{{Larson}(1981)}]{larson81}
{Larson}, R.~B. 1981, \mnras, 194, 809

\bibitem[{{Lee} {et~al.}(2015){Lee}, {Chang}, \& {Murray}}]{lee_etal15}
{Lee}, E.~J., {Chang}, P., \& {Murray}, N. 2015, \apj, 800, 49

\bibitem[{{Li} \& {Gnedin}(2014)}]{li_gnedin14}
{Li}, H., \& {Gnedin}, O.~Y. 2014, \apj, 796, 10

\bibitem[{{Mac Low} \& {Klessen}(2004)}]{maclow_klessen_04}
{Mac Low}, M.-M., \& {Klessen}, R.~S. 2004, Reviews of Modern Physics, 76, 125

\bibitem[{{McKee} \& {Ostriker}(2007)}]{mckee_ostriker07}
{McKee}, C.~F., \& {Ostriker}, E.~C. 2007, \araa, 45, 565

\bibitem[{{Muratov} \& {Gnedin}(2010)}]{muratov_gnedin10}
{Muratov}, A.~L., \& {Gnedin}, O.~Y. 2010, \apj, 718, 1266

\bibitem[{{Murray}(2011)}]{murray11}
{Murray}, N. 2011, \apj, 729, 133

\bibitem[{{Murray} \& {Chang}(2015)}]{murray_chang15}
{Murray}, N., \& {Chang}, P. 2015, \apj, 804, 44

\bibitem[{{Navarro} \& {White}(1993)}]{navarro_white_93}
{Navarro}, J.~F., \& {White}, S.~D.~M. 1993, \mnras, 265, 271

\bibitem[{{Paardekooper} {et~al.}(2015){Paardekooper}, {Khochfar}, \& {Dalla
  Vecchia}}]{paardekooper_etal15}
{Paardekooper}, J.-P., {Khochfar}, S., \& {Dalla Vecchia}, C. 2015, \mnras,
  451, 2544

\bibitem[{{Padoan} {et~al.}(2012){Padoan}, {Haugb{\o}lle}, \&
  {Nordlund}}]{padoan_etal12}
{Padoan}, P., {Haugb{\o}lle}, T., \& {Nordlund}, {\AA}. 2012, \apjl, 759, L27

\bibitem[{{Padoan} \& {Nordlund}(2011)}]{padoan_nordlund11}
{Padoan}, P., \& {Nordlund}, {\AA}. 2011, \apj, 730, 40

\bibitem[{{Planck Collaboration} {et~al.}(2014){Planck Collaboration}, {Ade},
  {Aghanim}, {Armitage-Caplan}, {Arnaud}, {Ashdown}, {Atrio-Barandela},
  {Aumont}, {Baccigalupi}, {Banday}, \& et~al.}]{planck_etal14}
{Planck Collaboration}, {Ade}, P.~A.~R., {Aghanim}, N., {et~al.} 2014, \aap,
  571, A15

\bibitem[{{Planck Collaboration} {et~al.}(2015){Planck Collaboration}, {Ade},
  {Aghanim}, {Arnaud}, {Ashdown}, {Aumont}, {Baccigalupi}, {Banday},
  {Barreiro}, {Bartlett}, \& et~al.}]{planck_etal15}
---. 2015, ArXiv e-prints, arXiv:1502.01589

\bibitem[{{Portegies Zwart} {et~al.}(2010){Portegies Zwart}, {McMillan}, \&
  {Gieles}}]{portegies_zwart_etal10}
{Portegies Zwart}, S.~F., {McMillan}, S.~L.~W., \& {Gieles}, M. 2010, \araa,
  48, 431

\bibitem[{{Read} {et~al.}(2015){Read}, {Agertz}, \& {Collins}}]{read_etal15}
{Read}, J.~I., {Agertz}, O., \& {Collins}, M.~L.~M. 2015, ArXiv e-prints,
  arXiv:1508.04143

\bibitem[{{Renaud} {et~al.}(2015){Renaud}, {Bournaud}, \&
  {Duc}}]{renaud_etal15}
{Renaud}, F., {Bournaud}, F., \& {Duc}, P.-A. 2015, \mnras, 446, 2038

\bibitem[{{Rudd} {et~al.}(2008){Rudd}, {Zentner}, \& {Kravtsov}}]{rudd_etal08}
{Rudd}, D.~H., {Zentner}, A.~R., \& {Kravtsov}, A.~V. 2008, \apj, 672, 19

\bibitem[{{Salem} \& {Bryan}(2014)}]{salem_bryan_14}
{Salem}, M., \& {Bryan}, G.~L. 2014, \mnras, 437, 3312

\bibitem[{{Scannapieco} {et~al.}(2008){Scannapieco}, {Tissera}, {White}, \&
  {Springel}}]{scannapieco_etal08}
{Scannapieco}, C., {Tissera}, P.~B., {White}, S.~D.~M., \& {Springel}, V. 2008,
  \mnras, 389, 1137

\bibitem[{{Schmidt} {et~al.}(2014){Schmidt}, {Almgren}, {Braun}, {Engels},
  {Niemeyer}, {Schulz}, {Mekuria}, {Aspden}, \& {Bell}}]{schmidt14}
{Schmidt}, W., {Almgren}, A.~S., {Braun}, H., {et~al.} 2014, \mnras, 440, 3051

\bibitem[{{Semenov} {et~al.}(2015){Semenov}, {Kravtsov}, \&
  {Gnedin}}]{semenov_etal15}
{Semenov}, V.~A., {Kravtsov}, A.~V., \& {Gnedin}, N.~Y. 2015, ArXiv e-prints,
  arXiv:1512.03101

\bibitem[{{Somerville} \& {Dav{\'e}}(2015)}]{somerville_dave15}
{Somerville}, R.~S., \& {Dav{\'e}}, R. 2015, \araa, 53, 51

\bibitem[{{Springel} {et~al.}(2006){Springel}, {Frenk}, \&
  {White}}]{springel_etal06}
{Springel}, V., {Frenk}, C.~S., \& {White}, S.~D.~M. 2006, \nat, 440, 1137

\bibitem[{{Stinson} {et~al.}(2006){Stinson}, {Seth}, {Katz}, {Wadsley},
  {Governato}, \& {Quinn}}]{stinson_etal06}
{Stinson}, G., {Seth}, A., {Katz}, N., {et~al.} 2006, \mnras, 373, 1074

\bibitem[{{Stinson} {et~al.}(2013){Stinson}, {Brook}, {Macci{\`o}}, {Wadsley},
  {Quinn}, \& {Couchman}}]{stinson_etal13}
{Stinson}, G.~S., {Brook}, C., {Macci{\`o}}, A.~V., {et~al.} 2013, \mnras, 428,
  129

\bibitem[{{Sun} {et~al.}(2016){Sun}, {de Grijs}, {Fan}, \&
  {Cameron}}]{sun_etal16}
{Sun}, W., {de Grijs}, R., {Fan}, Z., \& {Cameron}, E. 2016, \apj, 816, 9

\bibitem[{{Urquhart} {et~al.}(2014){Urquhart}, {Moore}, {Csengeri}, {Wyrowski},
  {Schuller}, {Hoare}, {Lumsden}, {Mottram}, {Thompson}, {Menten}, {Walmsley},
  {Bronfman}, {Pfalzner}, {K{\"o}nig}, \& {Wienen}}]{urquhart_etal14}
{Urquhart}, J.~S., {Moore}, T.~J.~T., {Csengeri}, T., {et~al.} 2014, \mnras,
  443, 1555

\bibitem[{{Vikhlinin} {et~al.}(2009){Vikhlinin}, {Kravtsov}, {Burenin},
  {Ebeling}, {Forman}, {Hornstrup}, {Jones}, {Murray}, {Nagai}, {Quintana}, \&
  {Voevodkin}}]{vikhlinin_etal09}
{Vikhlinin}, A., {Kravtsov}, A.~V., {Burenin}, R.~A., {et~al.} 2009, \apj, 692,
  1060

\bibitem[{{Vutisalchavakul} {et~al.}(2016){Vutisalchavakul}, {Evans}, \&
  {Heyer}}]{vutisalchavakul_etal16}
{Vutisalchavakul}, N., {Evans}, II, N.~J., \& {Heyer}, M. 2016, {\apj} in press
  (arXiv/1607.06518), arXiv:1607.06518

\bibitem[{{Wetzel} {et~al.}(2016){Wetzel}, {Hopkins}, {Kim}, {Faucher-Giguere},
  {Keres}, \& {Quataert}}]{wetzel_etal16}
{Wetzel}, A.~R., {Hopkins}, P.~F., {Kim}, J.-h., {et~al.} 2016, ArXiv e-prints,
  arXiv:1602.05957

\end{thebibliography}

\end{document}